\documentclass[fleqn,usenatbib]{mnras}
\usepackage{graphicx}
\usepackage{amsmath}
\usepackage{amssymb}
\usepackage{multicol}
\usepackage{bm}	
\usepackage{pdflscape}
\usepackage{comment}
\usepackage{subcaption}
\usepackage{mwe}
\usepackage{float}
\usepackage{caption}
\usepackage{amsmath}
\usepackage{mathtools}
\usepackage{soul}
\usepackage[mathscr]{eucal}
\usepackage{hyperref,tablefootnote,footnotehyper}
\usepackage[normalem]{ulem}
\hypersetup{colorlinks}
\usepackage[T1]{fontenc}
\usepackage{ae,aecompl}
\usepackage{newtxtext,newtxmath}

\title[Hercules Cluster in X-rays]{\hspace{0.5cm}The Hercules cluster in X-rays with \textit{XMM-Newton} and \textit{Chandra}}
\author[Tiwari $\&$ Singh]{\hspace{5.3cm} Juhi Tiwari\thanks{e-mail: tiwarijuhi92@gmail.com}
and Kulinder Pal Singh\thanks{e-mail: kpsinghx52@gmail.com}
\\\\
\hspace{2.7cm}Department of Physical Sciences, Indian Institute of Science Education and Research Mohali, Knowledge city,\\\hspace{5.1cm}Sector 81, Manauli, Sahibzada Ajit Singh Nagar, Punjab 140306, India
}
\begin{document}
\label{firstpage}
\pagerange{\pageref{firstpage}--\pageref{lastpage}}
\maketitle
\begin{abstract}
We present a detailed X-ray study of the central subcluster of the nearby ($z$ $\sim$0.0368) Hercules cluster (Abell 2151) identified as A2151C that shows a bimodal structure. A bright clump of hot gas with X-ray emission extending to radius $r$ $\sim$304 kpc and $L_X = 3.03_{-0.04}^{+0.02}\times10^{43}$ erg s$^{-1}$ in the 0.4$-$7.0 keV energy range is seen as a fairly regular subclump towards the west (A2151C(B)).  An irregular, fainter and cooler subclump with radius $r$ $\sim$364 kpc is seen towards the east (A2151C(F)) and has $L_X=1.13\pm{0.02}\times10^{43}$ erg s$^{-1}$ in the 0.4$-$7.0 keV energy band. The average temperature and elemental abundance of A2151C(B) are $2.01\pm{0.05}$ keV and $0.43\pm{0.05}$ Z$_{\odot}$ respectively, while these values are $1.17\pm{0.04}$ keV and $0.13\pm{0.02}$ Z$_{\odot}$ for A2151C(F). Low temperature (${1.55}\pm{0.07}$ keV) and a short cooling time ($\sim$0.81 Gyr) within the central 15 arcsec region confirm the presence of a cool core in A2151C(B). We identify several compact groups of galaxies within A2151C(F). We find that A2151C(F) is a distinct galaxy group in the process of formation and likely not a ram-pressure stripped part of the eastern subcluster in Hercules (A2151E). X-ray emission from A2151C shows a region of overlap  between A2151C(B) and A2151C(F) but without any enhancement of temperature or entropy in the two-dimensional (2D) projected thermodynamic maps that could have indicated an interaction due to merger between the two subclumps.
\end{abstract}

\section{Introduction} \label{sec:sec1}
Clusters of galaxies are the largest gravitationally collapsed concentrations of matter in the Universe and are believed to form through the continuous merging of smaller galaxy groups \citep{voit2005}. They are, therefore, very crucial for understanding the large-scale structure formation and evolution. Galaxy clusters are generally thought of as relaxed and evolved systems with total masses of $\sim$$10^{14}-10^{15}$ M$_{\odot}$ \citep{mcnamara2007,kravtsov2012} and large sizes of several Mpc. Observations of clusters, however, reveal that they contain a significant degree of substructure, indicating that they are dynamically evolving systems \citep{fabian1992,oergerle2001}. The substructures can be seen in the galaxy surface number density distribution \citep{geller1982,flin2006} and X-ray surface brightness (SB) maps of clusters \citep{gomez1997,jones1999,schuecker2001} in the form of multiple peaks. The merging of substructures into larger cluster units is often visible as a sudden rise in the intra-cluster medium (ICM) temperature and entropy, or as a discontinuity in the X-ray SB profiles \citep{markevitch2007,bohringer2010}.

Substructures within clusters usually coincide with smaller groups of galaxies which are known to have gas properties that are different from their larger counterparts \citep{mulchaey2000,lovisari2015}. Galaxy groups are not simple scaled-down versions of the bigger clusters. A number of X-ray studies of the ICM of different cluster samples \citep{ponman1999,helsdon2000,maughan2012} suggest that low temperature systems ($< 3.5$ keV; mainly galaxy groups) show a significant departure from simple power-law relationships like $L_{X}\propto (kT)^{2}$ where $L_{X}$ and $kT$ are the X-ray gas luminosity and temperature, respectively. These power-law relations are predicted for clusters under the assumptions that they form via a single gravitational collapse and the only source of energy input into the ICM is gravitational \citep{kaiser1986}.   Departure from power-law relations is believed to be a consequence of non-gravitational processes (cooling, AGN feedback, star formation, and galactic winds) which play a very important role in the formation and evolution of galaxy groups due to their shallow potential wells \citep{eckmiller2011,lovisari2015}. The study of individual substructures is, therefore, crucial to understand the role of these processes in cluster formation.

Hercules cluster of galaxies (Abell 2151 -- hereafter A2151) at a redshift, $z$, of 0.0368 \citep{zabludoff93} is a significantly substructured system. Its optical centre as determined by \citet{agulli2017} is at $\text{R.A. (J2000)}=16^{h}05^{m}23^{s}.369$; $\text{Dec. (J2000)}=17^{\circ} 45' 04''.33$, and it has $\sim$360 galaxies belonging to it within  a radius of $\sim$1.3$\text{R}_{200}$ \footnote{R$_{200}$ is the radius from the adopted cluster centre at which the density is 200 times the critical density of the universe; It is equal to 1.45 Mpc for A2151 as obtained by \citet{agulli2017}}. The cluster contains a high fraction of spiral galaxies ($\gtrsim50$ per cent; \citet{gioandhaynes85,bdb95,maccagni95}) and has a clumpy gas distribution \citep{magri88,bdb95}. A2151, along with A2147 and A2152, is part of a much larger Hercules supercluster \citep{barmby1998}.

\begin{figure*}
\centering
   \begin{subfigure}[b]{\textwidth}
   \makebox[\textwidth][c]{\includegraphics[width=25cm,height=11cm]{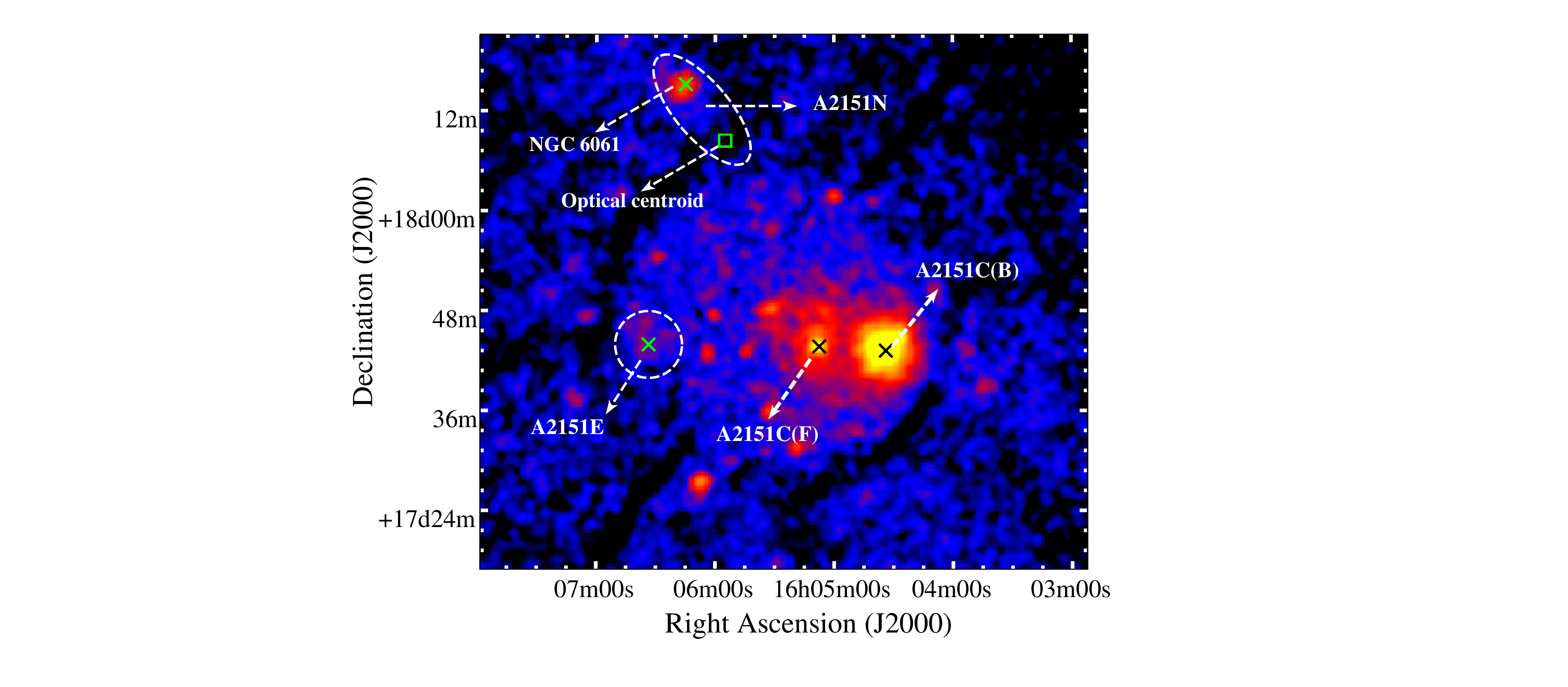}}
\end{subfigure}
\begin{subfigure}[b]{\textwidth}
   \makebox[\textwidth][c]{\includegraphics[width=16cm,height=10cm]{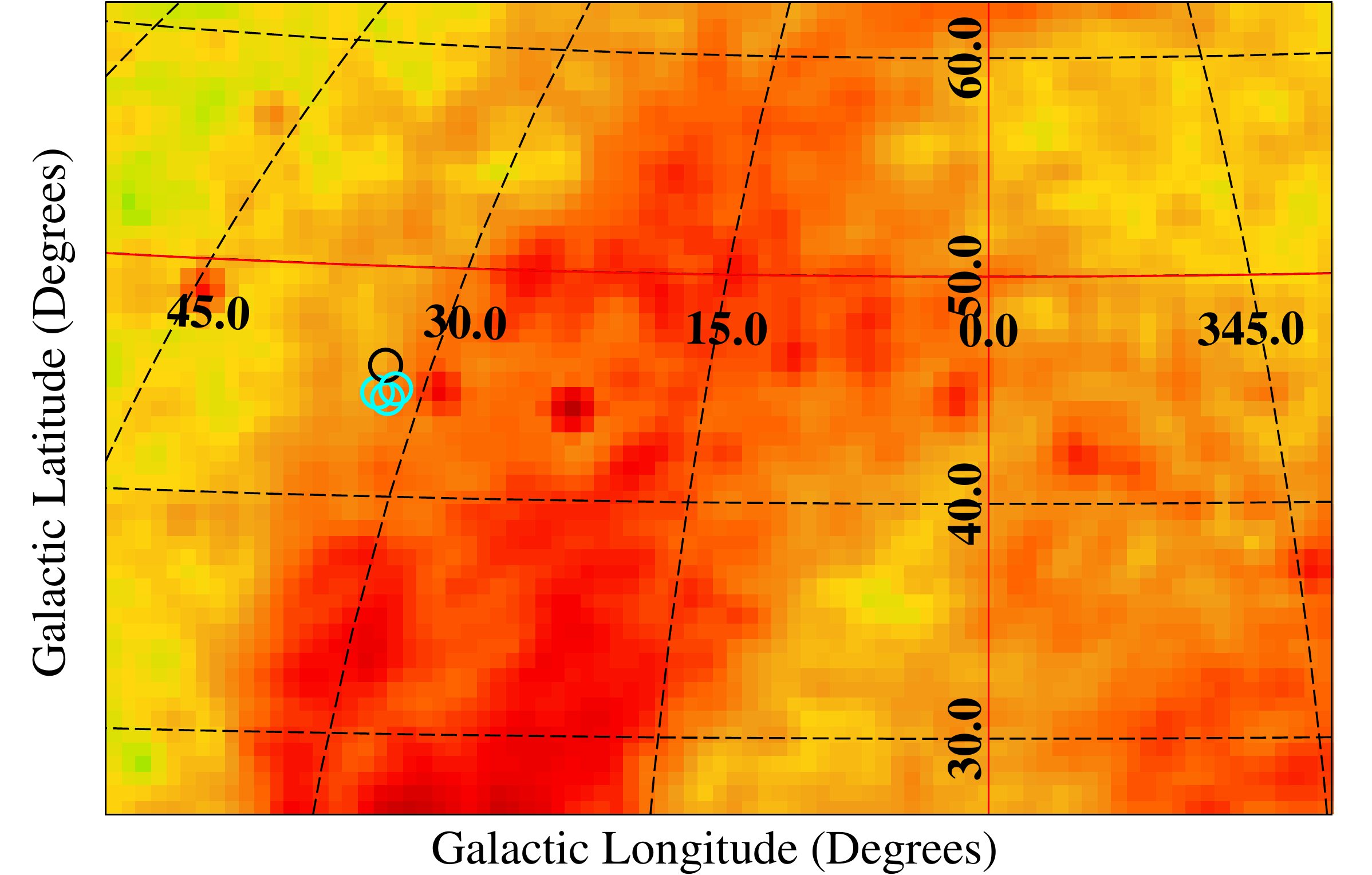}}
\end{subfigure}
\vspace{-1\baselineskip}
\caption{\textit{\textbf{Top panel}}: \textit{ROSAT} PSPC image of A2151 showing the different cluster components -- A2151C, A2151E, and A2151N. The image was generated using data from the HEASARC archive. The crosses mark the positions of the \textit{ROSAT} X-ray peaks obtained for each subcluster by \citet{bdb95}. A2151C is the brightest and has a two-component structure. A2151E towards the east is fainter. A2151N has negligible X-ray emission associated with cluster gas \citep{bdb95}. The primary source of X-ray emission in A2151N is the galaxy NGC 6061. The green box symbol marks the position of the optical centroid of A2151N; \textit{\textbf{Bottom panel}}: RASS $3/4$ keV diffuse background image showing part of the North Polar Spur (NPS)\textsuperscript{a}. The cyan circles mark the positions of the three Hercules subclusters -- A2151C, A2151N, and A2151E. The black circle in their vicinity indicates the position of the region used to obtain the RASS diffuse background spectrum. The circles are only for representation and do not indicate the actual sizes of the subclusters/background extraction region.}
\label{fig:fig0}
\small\textsuperscript{a}The NASA SkyView service was used to generate the image.\\\url{https://skyview.gsfc.nasa.gov/current/cgi/query.pl}
\end{figure*}

A2151 has previously been studied in X-rays by \citet{bdb95} using data from the Position Sensitive Proportional Counter (PSPC) aboard the Roentgen Satellite (\textit{ROSAT}), and by \citet{huangsarazin96} who used \textit{ROSAT}'s High Resolution Imager (HRI). The PSPC has a field of view (fov) $\sim$2\textdegree{} across while the HRI has a fov $\sim$38 arcmin across. The PSPC image of \citet{bdb95} (fig. 1 of their paper) revealed the presence of three subclusters in A2151 -- a bright central bimodal gas clump (A2151C), a fainter eastern clump (A2151E) and a northern clump (A2151N) with negligible X-ray emission. The HRI image confirmed the two component X-ray structure of A2151C -- a bright central subclump, A2151C(B), and a faint subclump, A2151(F). These subclumps were noticed earlier in an X-ray image observed with the Einstein Observatory's Imaging Proportional Counter (IPC) \citep{magri88}. \textit{ROSAT} PSPC image of A2151 showing its different components is displayed in \autoref{fig:fig0}\textit{ (top panel)}.

A global X-ray spectral analysis of A2151C(B) (for a radius (\textit{r}) of $\text{210 } h^{-1}_{75} \text{ kpc}$ from the X-ray peak), A2151C(F) ($\textit{r}=\text{200 } h^{-1}_{75} \text{ kpc}$), A2151E ($\textit{r}=\text{130 } h^{-1}_{75} \text{ kpc}$) and A2151N ($\textit{r}=\text{151 } h^{-1}_{75} \text{ kpc}$) was performed by \citet{bdb95} in the energy band 0.1-2.4 keV using low spectral resolution of \textit{ROSAT} PSPC ($\sim$550 eV at 2 keV).  They found that A2151C(B) and A2151C(F) have X-ray temperatures of $1.67^{+0.47}_{-0.25}\text{ keV} \text{ and } 1.03^{+0.60}_{-0.06}$ keV, and overall elemental abundances of $0.56^{+0.31}_{-0.20} \text{ and } 0.32^{+0.15}_{-0.10}$ times solar, respectively. The X-ray luminosities of A2151C(B) and A2151C(F) in the 0.1-2.0 keV band were estimated to be $8.7\pm{0.7} \times 10^{42} \text{ erg  s}^{-1}$ and $3.6^{+0.68}_{-0.61} \times 10^{42}$ erg s$^{-1} $ respectively. A spectral analysis of different regions within the individual subclusters was not carried out by \citet{bdb95}. The surface brightness analysis of A2151C(B) by \citet{huangsarazin96} with the \textit{ROSAT} HRI image revealed the presence of central excess X-ray emission in the subclump indicating that it contains a cooling flow. The authors estimated a central cooling time of $< 1.9 \times 10^{8} \text{ yr}$ within the central 4 arcsec radius of A2151C(B).

Evidence for subclustering in A2151 was also seen in the radial velocity data of 127 galaxies compiled by  \citet{bds93} (within 0\textdegree{}.9 radius centered at $\text{R.A.(J2000)}=16^{h}03^{m}11^{s}$;  $\text{Dec.(J2000)}=17^{\circ} 55' 55''$). They had identified the presence of three galaxy groups or subclusters in their galaxy surface density map.  These subclusters were identified as dynamically independent systems by both \citet{bds93} and \citet{bdb95}, and coincided with the three X-ray gas clumps. The presence of these three substructures was also seen in the wavelet analysis image of \citet{escalera94} whose galaxy sample consisted of 79 galaxies. \citet{huangsarazin96} used the galaxy list of \citet{bds93} and found another galaxy group A2151S to the south of A2151C in their luminosity-weighted galaxy surface density map in addition to the three previously identified subclusters. Moreover, their galaxy density distribution map (fig. 2 of \citet{huangsarazin96}) showed that the central subcluster further divided into two galaxy groups which coincided with the two components (A2151C(B) and A2151C(F)) seen in the X-ray emission of A2151C. A more recent spectroscopic study of A2151 by \citet{agulli2017} using data from the William Herschel Telescope (WHT) has confirmed the presence of substructure within A2151C (\S3.2 and fig. 5 of their paper),  identifying 20 members within  A2151C(B) and 57 members in A2151C(F). The mean velocity and velocity dispersion were estimated to be 10116 km $\text{s}^{-1}$ and 441 km $\text{s}^{-1}$ respectively for A2151C(B), and 10299 km $\text{s}^{-1}$ and 711 km $\text{s}^{-1}$ respectively for A2151C(F).

Here, we present a study of hot X-ray emitting gas within the central subcluster of Hercules (A2151C) using all  available archival data from observations with the European Photon Imaging Camera (EPIC) aboard the \textit{XMM-Newton} observatory and the Advanced CCD Imaging Spectrometer (ACIS) aboard the \textit{Chandra} X-ray Observatory (CXO).  This study exploits the superior  spatial and energy resolution of \textit{XMM-Newton} ($\sim$5 arcsec; $\sim$120 eV) and \textit{Chandra} ($\sim$0.5 arcsec; $\sim$120 eV) over the \textit{ROSAT} PSPC ($\sim$25 arcsec, $\sim$550 eV) and \textit{ROSAT} HRI which had a good spatial resolution ($\sim$2 arcsec) but no spectral resolution.   A detailed X-ray spectral analysis of A2151C was, therefore, not possible with \textit{ROSAT}.    Besides, the \textit{ROSAT} was sensitive only to soft X-rays between 0.1 to 2.0 keV in contrast to the wider energy bandwidth of \textit{Chandra} ($\sim$0.2$-$10.0 keV) and \textit{XMM-Newton} ($\sim$0.2$-$10.0 keV).   We use the capabilities of \textit{Chandra} and \textit{XMM-Newton} to investigate the thermodynamic properties of the ICM of A2151C by performing a detailed spectral analysis within the two substructures, A2151C(B) and A2151C(F), which are now optically associated with two different galaxy groups \citep{huangsarazin96,agulli2017}.   We obtain radial profiles of the properties of the gas in the two groups, and 2D projected maps of metallicity and thermodynamic (TD) properties of the gas in A2151C.

The paper is organized as follows. The observations used here and details of the data reduction are provided in \S2. The imaging and spectral analyses along with the radial profiles of the properties of the gas are presented in \S3. Estimates of the cooling time, gas mass and total mass of the two groups are also given in this section. A discussion based on the results is presented in \S4.  Finally, we present our conclusions and results in \S5.  We use H$_{0}$ = 67.4 km s$^{-1}$ Mpc$^{-1}$, $\Omega_{m}$ = 0.315 and $\Omega_{\Lambda}$ = 0.685 based on the findings of \citet{planck2018}, in all our calculations. 
At redshift $z$ = 0.0368 corresponding to A2151, 1 arcsec equals 0.76 kpc \citep{wright2006}.

\section{Observations and Data Reduction}
All the publicly available \textit{XMM-Newton} and \textit{Chandra} X-ray observations of A2151 are listed in \autoref{tab:tab1}. X-ray data from these observations were obtained from the HEASARC\footnote{High Energy Astrophysics Science Archive Research Center\\  \url{https://heasarc.gsfc.nasa.gov/}} archive. Not all of these observations were used in the imaging and spectral analyses. The reasons for including/discarding a particular observation are discussed below.

\begin{table*}
\centering
\captionsetup{justification=centering}
  \caption{\textit{XMM-Newton} and \textit{Chandra} X-ray Observations of A2151}
  \begin{tabular}{cccccccc}
    \hline
\hline
Satellite & Observation ID & Date of Obs. &RA & Dec & Detector & Useful Exposure time & Observation time \\
&&&(J2000)&(J2000)&&(ks)&(ks)\\
&&&h:min:s&\textdegree{} : $'$ : $''$&&&\\
\hline
\hline
\textit{XMM-Newton}&0147210201\textsuperscript{a}&2003 March 10&16 05 07.08&+17 44 59.00&PN&0.72&26.20\\
&&&&&MOS1&0.84&\\
&&&&&MOS2&1.68&\\
&0147210101\textsuperscript{a}&2003 Aug 9&16 05 07.08&+17 44 59.00&PN&1.56&29.95\\
&&&&&MOS1&3.54&\\
&&&&&MOS2&3.72&\\
&0147210301&2003 Aug 11&16 05 07.08&+17 44 59.00&PN&4.98&19.04\\
&&&&&MOS1&7.74&\\
&&&&&MOS2&8.16&\\
\hline
\textit{Chandra}&4996\textsuperscript{b}&2004 July 25&16 04 36.00&+17 43 23.00&ACIS-I&14.22&22.12\\
&17038\textsuperscript{c}&2015 Feb 22&16 05 31.80&+17 48 26.20&ACIS-S&7.85&7.85\\
&18171\textsuperscript{c}&2016 April 17&16 05 44.60&+17 43 03.50&ACIS-S&15.07&15.07\\
&19592&2017 May 31&16 04 35.80&+17 43 17.30&ACIS-I&16.46&16.46\\
&20086&2017 June 03&16 04 35.80&+17 43 17.30&ACIS-I&30.06&30.06\\
&20087&2017 June 04&16 04 35.80&+17 43 17.30&ACIS-I&35.57&35.57\\
\hline\hline
  \end{tabular}
\label{tab:tab1}
\\\small\textsuperscript{a}These observations were strongly affected by SP flares and were excluded from all imaging and spectral analyses;
\\\small\textsuperscript{b}The observation was excluded from all imaging and spectral analyses due to strong residual SP contamination even after a strict LC filtering.
\\\small\textsuperscript{c}These observations were used only in the imaging analysis.
 \end{table*}
 
\begin{figure*}
\centering
\begin{tabular}{cc}
\includegraphics[width=0.53\textwidth,height=5.85cm]{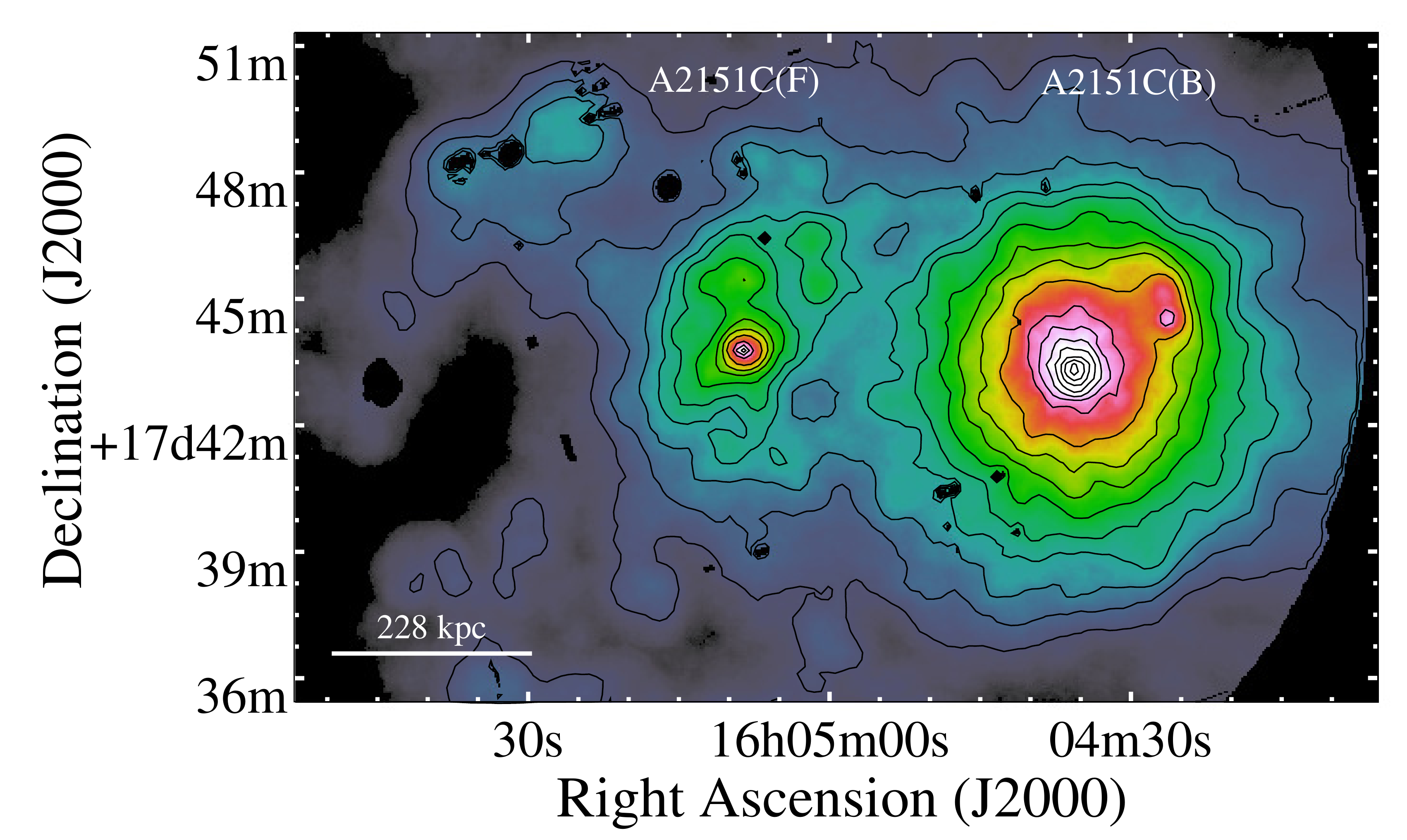} &
\includegraphics[width=0.5\textwidth]{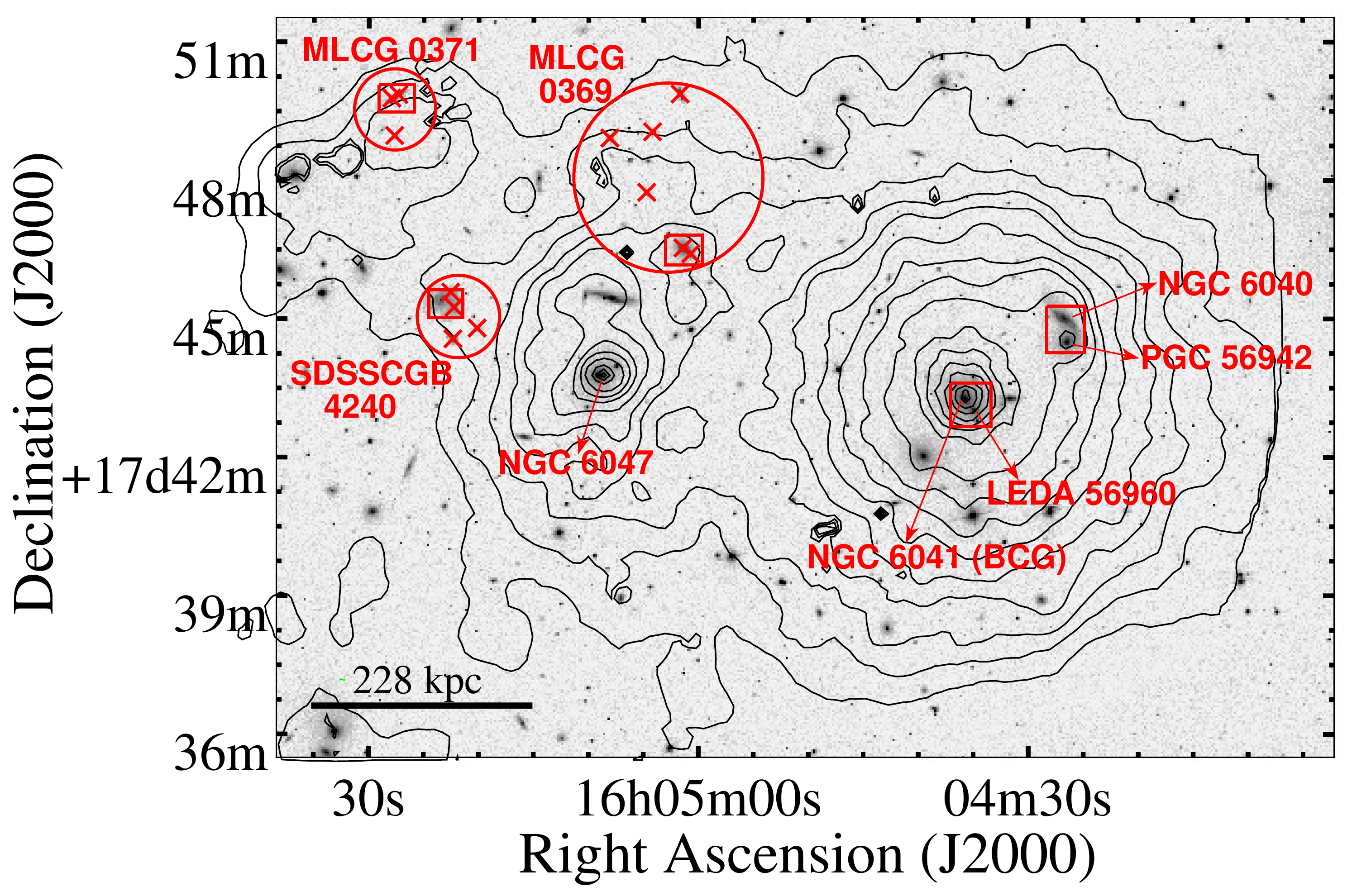} \\
\hspace{1.6cm}\textbf{(a)} & \hspace{1.7cm}\textbf{(b)}
\end{tabular}
\begin{tabular}{ccc}
&\includegraphics[width=0.525\textwidth,height=5.7cm]{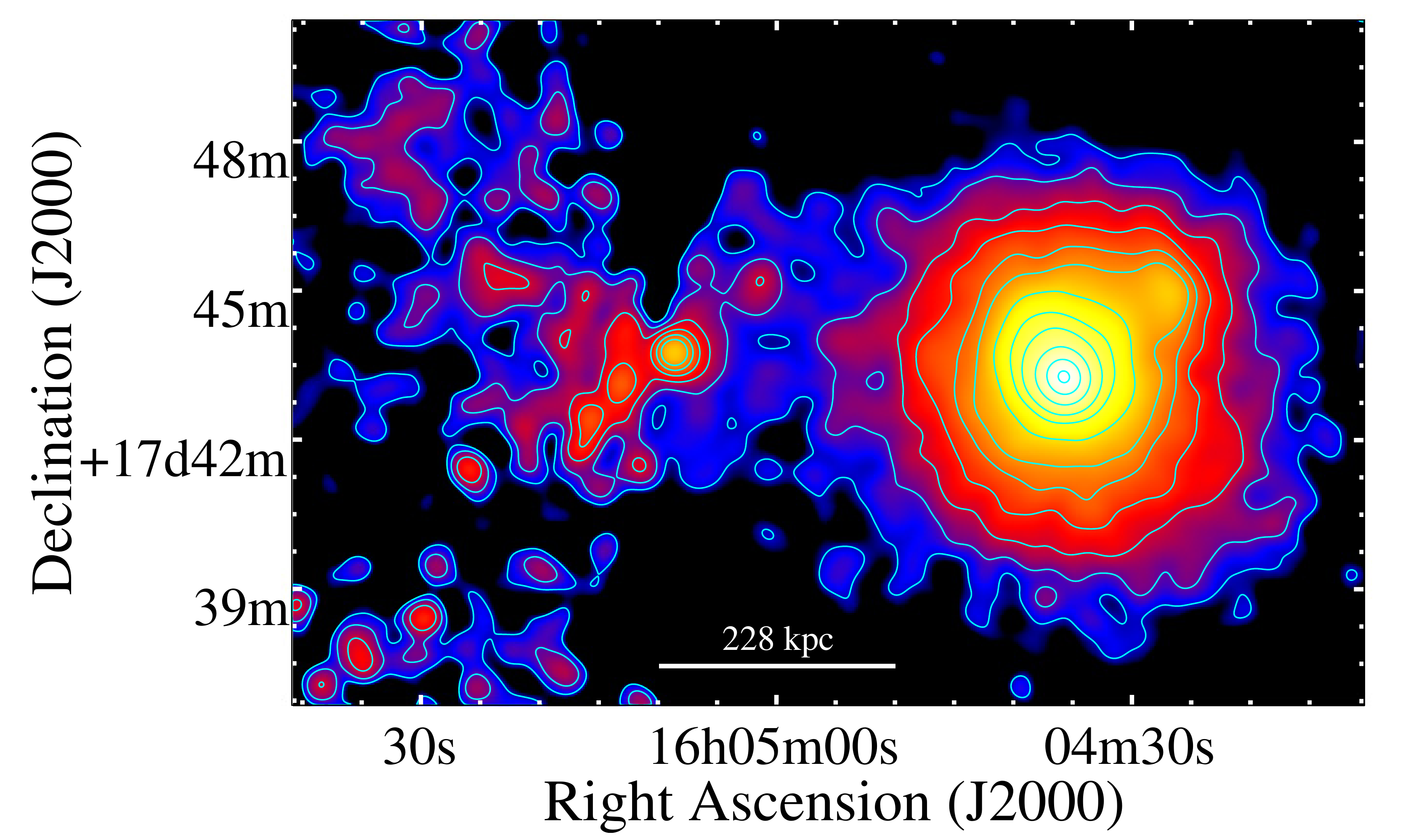}& \\
&\hspace{2cm}\textbf{(c)}&
\end{tabular}
\caption{ \textbf{(a)}: The combined PN, MOS1, and MOS2 \textit{XMM-Newton} image in the 0.3--4.0 keV energy band created after particle-background subtraction and exposure correction. The image was adaptively smoothed using 200 counts. Both A2151C(B) (bright western subclump) and A2151C(F) (faint eastern subclump) are visible due to the 0\textdegree{}.5 diameter fov of \textit{XMM-Newton}. Overlaid X-ray surface brightness contours are logarithmically distributed between 9.6 x 10$^{-7}$ and $8.5 \text{ x } 10^{-5} \text{cts s}^{-1}$ arcsec$^{-2}$. \textbf{(b)}: The optical r-band image from the Sloan Digital Sky Survey with overlaid X-ray contours from the \textit{XMM-Newton} image shown in panel (a). Five pairs of interacting galaxies and three compact galaxy groups within A2151C identified from NED\textsuperscript{a} and SIMBAD\textsuperscript{b} databases are highlighted in red boxes and red circles respectively. The circles are used only for representation and do not indicate the actual angular size of the group. The red crosses represent the galaxies identified within the individual compact groups.  \textbf{(c)}: The 0.4--4.0 keV exposure-corrected mosaic created using the five \textit{Chandra} observations after elimination of point sources and subtraction of particle and X-ray background. The image was smoothed using a gaussian kernel with $\sigma$ = 15 arcsec. Overlaid logarithmic contours range from 1.1 x 10$^{-10}$ to 4.0 x 10$^{-8}$ photons cm$^{-2}$ s$^{-1}$ pixel$^{-1}$.}
\label{fig:fig1}
\small\textsuperscript{a}NASA/IPAC Extragalactic Database: \url{http://ned.ipac.caltech.edu/};
\\\small\textsuperscript{b}Set of Identifications, Measurements and Bibliography for Astronomical Data: \url{http://simbad.u-strasbg.fr/} 
\end{figure*}

\subsection{XMM-Newton}\label{sec2.1}
\textit{\textbf{Data Screening:}} The \textit{XMM-Newton} Extended Source Analysis Software (XMM-ESAS) package which is integrated into SAS (version 17.0.0) was used for all data reduction. The raw data of each observation were processed using the SAS tasks \textit{epchain} and \textit{emchain} along with the latest calibration files to produce the PN and MOS event files respectively. 
These event files were then filtered for Soft Proton (SP) flares using the tasks \textit{pn-filter} and \textit{mos-filter}. The useful exposure time obtained as a result of this filtering is given in \autoref{tab:tab1}. The observations 0147210101 and 0147210201 were strongly contaminated with SP flares and had very low useful exposure time values. These two observations were, therefore, excluded from all further analyses. Out of the three available observations of A2151, only 0147210301 was found to be useful. We have used the EPIC PN, MOS1 and MOS2 data of this observation in this work. During the observation the PN detector was operated in Extended Full Frame mode with thin filter while the MOS detectors were set to Full Frame mode and used thin filter. None of the CCDs in the MOS detectors was found to be in an anomalous state for 0147210301.

\textbf{\textit{Point source detection:}} The XMM-ESAS metatask \textit{cheese} was used to run point source detection on full-field images (it calls the SAS task \textit{edetect$\_$chain}) and create cheese masks from the output point source list. A detection box of size 5$\times{}5$ pixels was used. The task \textit{edetect$\_$chain} also calculates individual exposure maps for the three EPIC detectors.

\textbf{\textit{Quiescent Particle Background treatment:}} The XMM-ESAS package allows neat modelling of the quiescent particle background (QPB) using a combination of the filter-wheel-closed (FWC) data and a database of unexposed-region data based on the methods of \citet{kuntz2008}. No blanksky files were used in our analysis.

\textbf{\textit{X-ray background treatment:}} The X-ray background (XRB) can be separated into two main components. One is the cosmic X-ray background (CXB) which is mainly attributed to point sources of extragalactic origin, mostly believed to be Active Galactic Nuclei (AGNs). The CXB dominates the X-ray background in the energy band 2.0--10.0 keV and is well fit by a non-thermal power-law with photon index of $\sim$1.4 \citep{lumb2002,cxb2004,hickox2006,cappelluti2017}. The second component is the soft emission from within our Milky Way Galaxy. It is the dominant contributor to the XRB at energies below 2.0 keV and is strongly dependent on the direction of observation. We refer to it as the soft X-ray background (SXRB) here. The SXRB mainly consists of three components -- an unabsorbed $\sim$0.1 keV thermal emission from the Local Hot Bubble (LHB), an absorbed $\sim$0.1 keV thermal emission from the Galactic Halo (GH), and a third hotter ($\sim $0.26 keV) absorbed thermal component associated with the North Polar Spur (NPS) \citep{willingale2003,snowden2008}. A2151 lies in the direction of the NPS as seen in \autoref{fig:fig0}\textit{ (bottom panel)} that shows the \textit{ROSAT} All Sky Survey (RASS) NPS image in the 3/4 keV band.  Therefore, it is very important to take into account the contribution from the SXRB in the spectral modelling. We have included the modelling of the complete X-ray background (CXB+SXRB) in the spectral analysis (\S3.2). In order to constrain the XRB parameters, we have used spectral data from the RASS. The RASS diffuse background spectrum and PSPC response were obtained for a circular region of radius 0\textdegree{}.6 (centred on galactic coordinates: l$^{II}$=32\textdegree{}.47, b$^{II}$=45\textdegree{}.66) in the vicinity of A2151 using the HEASARC X-ray background tool\footnote{\url{https://heasarc.gsfc.nasa.gov/cgi-bin/Tools/xraybg/xraybg.pl}}. The neutral and total hydrogen column densities within this region are similar to those in the A2151 region. This seven channel RASS background spectrum was simultaneously fitted, after proper correction for the observed solid angle, with the \textit{XMM-Newton} data by a standard model for the XRB (see \S3.2 for details of spectral modelling) \citep{snowden2008}.

\textbf{\textit{Mosaic making and spectral extraction:}} The task \textit{pn$\_$spectra} (\textit{mos$\_$spectra}) followed by \textit{pn$\_$back} (\textit{mos$\_$back}), with parameters \textit{elow=300 and ehigh=4000} and leaving the \textit{region} parameter empty, was used to generate the 0.3$-$4.0 keV PN (MOS1/MOS2) image of the entire field of view while masking out the detected point sources (parameter \textit{mask=1}). Running these tasks for the entire fov also generated PN, MOS1, and MOS2 model QPB images.  The intermediate merged files of the point-source-subtracted observed counts images, the exposure maps and the model particle background count images in the 0.3$-$4.0 keV band, were produced using the ESAS task \textit{comb}.  Finally, the task \textit{adapt} was used to generate the combined PN, MOS1, and MOS2, particle-background-subtracted, exposure-corrected image of A2151 in the energy range 0.3$-$4.0 keV. The merged image was smoothed using 200 counts and a pixel size of 2.5 arcsec was used.

The same ESAS tasks \textit{pn$\_$spectra} (\textit{mos$\_$spectra}) and \textit{pn$\_$back} (\textit{mos$\_$back}) were used to generate the source and model background spectra as well the EPIC responses.   In cases where spectra were desired from a particular region on the detector, a region expression in detector coordinates was input into the tasks \textit{pn$\_$spectra} and \textit{mos$\_$spectra}. All spectra were suitably grouped using the FTOOL \textit{grppha} (HEASoft version 6.27), and analysed using XSPEC version 12.11.0 \citep{xspec1996}. We note that the tasks \textit{pn$\_$back} and \textit{mos$\_$back} generate only the model QPB spectra. Therefore, the instrumental background lines, primarily the Al K$\alpha$ ($\sim$1.49 keV) and Si K$\alpha$ ($\sim$1.75 keV) for MOS and Al K$\alpha$ for PN, were explicitly added to the models to fit the data.

\subsection{Chandra X-ray Observatory}\label{sec2.2}
\textit{\textbf{Data Screening: }}
The data were analysed with Chandra Interactive Analysis of Observations (CIAO) software version 4.12 and CALDB version 4.9.1. Each dataset was reprocessed with the most recently available calibration applied to it using the \textit{chandra$\_$repro} script. The reprocessing created a new \textit{level=2} event file for each dataset which was used in all further processing. The \textit{level=2} event files were then filtered for SP flares by analysing their light curves (LCs) generated in the energy band 0.4$-$7.0 keV. The LCs of all observations except 4996 were found to be flare free. The good time intervals (GTIs) obtained after the LC filtering are listed in \autoref{tab:tab1}. The observation 4996, strongly affected by SP flares, was found to have residual SP contamination even after a 2.5$\sigma$ LC filtering using the CIAO task \textit{lc$\_$sigma$\_$clip} that resulted in a useful exposure time value of 14.22 ks for this observation. This residual contamination was found to adversely affect the results of the spectral fits, and therefore, 4996 was excluded from all further analyses. The observations -- 17038 and 18171 -- were found to be useful for the imaging analysis only. The observation 17038 covered the eastern part of the cluster only partially. Therefore, an average or radial spectral analysis of the eastern region with 17038 was not possible. The observation 18171 covered only the far eastern part of the cluster and beyond and was, therefore, not helpful in the spectral analysis. Thus, we have used five \textit{Chandra} observations of A2151, viz., 19592, 20086, 20087, 17038, and 18171 for the imaging analysis and three observations $-$ 19592, 20086, and 20087 for the spectral analysis.

\textbf{\textit{Particle Background treatment:}} Chandra allows the generation of a blanksky background file compatible with a given events file using the task \textit{blanksky}. The task makes use of the instrument-specific background files present in the CALDB which have to be specifically installed. Flare filtering but no energy filtering must be done in order to generate these blanksky files. We created the matching blanksky background file for each dataset using this task \textit{blanksky}.

\textbf{\textit{X-ray background treatment:}} The XRB treatment is the same as that described for the \textit{XMM-Newton} data in \S2.1 .

\textbf{\textit{Image generation and point source subtraction:}} The flare-free event files were filtered for the energy range 0.4$-$4.0 keV using the task \textit{dmcopy} in order to create images in the same energy band. The task \textit{fluximage} was then used to generate the counts image (0.4$-$4.0 keV), the exposure map at an effective energy of $2.3$ keV, and a psfmap at the same effective energy with an enclosed-count fraction (ECF) of 0.9 for each dataset. The pixel size used was 0.492 arcsec. The source counts image generated here included point sources which required detection and consequent subtraction.
 We used the task \textit{wavdetect} with wavelet \textit{scales=2,4} and \textit{ellsigma=5} to identify point sources in each dataset. A background region around each detected point source was required for filling the holes created in the point-source-subtracted image and was estimated using the task \textit{roi}. The task \textit{dmfilth} was then used to create a point-source-subtracted, hole-filled image for each dataset. Scaled background images were created for each observation given the observation-specific blanksky background files using the task \textit{blanksky$\_$image}.
 
\textbf{\textit{Mosaic generation:}} The point-source-subtracted, hole-filled, unsmoothed images of the three datasets were projected to a common centre and combined using \textit{reproject$\_$image$\_$grid}. We used a pixel size of 0.492 arcsec and a region of 6000$\times{}$6000 pixels. The individual background images and exposure maps were combined and reprojected according to the combined source image using the task \textit{reproject$\_$image}. Background subtraction and exposure correction of the combined source image was done using the task \textit{dmimgcalc} and the resulting mosaic was smoothed with \textit{aconvolve}.

\textbf{\textit{Spectral extraction:}} The point-source-subtracted event files were used for extraction of spectra and responses with the CIAO task \textit{specextract}. Background spectra were extracted from the scaled blanksky background events files generated using the \textit{blanksky$\_$sample} script. The spectra were appropriately grouped using the FTOOL \textit{grppha} and analysed using XSPEC version 12.11.0 .

\subsection{Sloan Digital Sky Survey}\label{sec2.3}
We have used the Sloan Digital Sky Survey (SDSS) data release 12 (DR12) imaging data to generate an optical mosaic of A2151. Forty two \textit{r-band} images were retrieved from the DR12 Science Archive Server in a field of 18 arcmin centred at RA = 241\textdegree{}.149 and Dec = +17\textdegree{}.72. These images were mosaicked into a single image using SWarp version 2.38.0.

\section{Imaging and Spectral analyses}\label{sec3}
\subsection{X-ray Morphology}\label{sec3.1}
An X-ray image of the central subcluster of Hercules, A2151C,  in the 0.3--4.0 keV energy band is shown in
\autoref{fig:fig1}\textcolor{blue}{(a)}.  The image has been made using combined data from PN+MOS1+MOS2 \textit{XMM-Newton} and has been corrected for exposures and has the particle-background removed. We estimated the average XRB model surface brightness in the energy range 0.3--4.0 keV from the global spectral analysis described in \S3.2.1. This resulted in XRB surface brightness values of $8.6\times 10^{-7}$, $2.1\times 10^{-7}$, and $2.8$ $\times 10^{-7}$ cts s$^{-1}$ arcsec$^{-2}$ for the PN, MOS1 and MOS2 detectors respectively. The value obtained for PN is naturally higher due to the higher efficiency of the detector. The XRB surface brightness contribution thus obtained has, however, not been subtracted from the combined image, since it was not possible to account for the variation in the efficiency of the three detectors. Nonetheless, these values indicate that the contribution of the XRB to the observed X-ray surface brightness is low. We note that \textit{XMM-Newton} combined PN+MOS1+MOS2 image takes the average of the contribution from the three detectors and also corrects for the variation in the detector efficiencies.

Two distinct X-ray clumps are seen in the image shown in \autoref{fig:fig1}\textcolor{blue}{(a)}. The presence of these two X-ray blobs was first reported in the Einstein IPC image by \cite{magri88}. They were also identified in the 0.1--2.4 keV \textit{ROSAT} PSPC image of \citet{bdb95} and the \textit{ROSAT} HRI image of \citet{huangsarazin96} who confirmed that the X-ray emission in the A2151C divides into two components which correspond to two groups of galaxies. Following the nomenclature of \citet{huangsarazin96}, we refer to the \textit{brighter western} and \textit{fainter eastern} X-ray clumps within A2151C as A2151C(B) (central bright) [X-ray peak: RA = 16\textsuperscript{h}04\textsuperscript{m}35\textsuperscript{s}.73 ; Dec = +17\textdegree{}43$'$17$''$.31] and A2151C(F) (central faint) [X-ray peak: RA = 16\textsuperscript{h}05\textsuperscript{m}08\textsuperscript{s}.70 ; Dec = +17\textdegree{}43$'$47$''$.96] respectively. The central peak of X-ray emission in A2151C(B) coincides with the brightest cluster galaxy (BCG) NGC 6041 (SIMBAD) while that in A2151C(F) is coincident with a radio galaxy, NGC 6047 (Fig. \ref{fig:fig1}\textcolor{blue}{(b)}). Both these galaxies have been classified as giant ellipticals by \citet{efigi2011}. NGC 6047 has also been designated as a BCG by \citet{ngc60472010pos}. An optical r-band image of A2151C from the Sloan Digital Sky Survey is displayed in \autoref{fig:fig1}\textcolor{blue}{(b)}, and it shows the clustering of galaxies around the two X-ray peaks. The division of A2151C into two distinct substructures has been observed earlier in the galaxy surface density distribution of \citet{huangsarazin96} and was confirmed by a more recent optical study by \citet{agulli2017}. Fig. \ref{fig:fig1}\textcolor{blue}{(b)} also shows several pairs of interacting galaxies within A2151C. A few compact galaxy groups (CGGs) have also been identified in the fainter subclump A2151C(F).

In \autoref{fig:fig1}\textcolor{blue}{(c)} we show the background-subtracted, exposure-corrected, 0.4--4.0 keV combined \textit{Chandra} image of A2151C generated using the five ACIS observations. The image covers A2151C(F) only partially. We note that in addition to the particle-background, the contribution of the XRB has also been subtracted from the mosaic. The 0.4--4.0 keV average surface brightness of the XRB was estimated from the spectral analysis described in \S3.2.2 and was found to be equal to $3.51\times10^{-10}$ photons cm$^{-2}$ s$^{-1}$ pixel$^{-1}$ (pixel size$=$0.492 arcsec). The observation 18171 was included in the imaging analysis for completeness. However, this observation has not contributed much to the image.

It is evident from both Fig. \ref{fig:fig1}\textcolor{blue}{(a)} and Fig. \ref{fig:fig1}\textcolor{blue}{(c)} that except for a notable departure from circular symmetry in the northwest region, X-ray emission is fairly azimuthally symmetric in A2151C(B). This northwest bump in X-ray emission overlaps with a pair of interacting galaxies NGC 6040 and PGC 56942 (identified as Arp 122 by \citet{arp1966} in his \textit{Atlas of Peculiar Galaxies}; see Fig. \ref{fig:fig1}\textcolor{blue}{(b)}), which are known to be substantially depleted of their neutral hydrogen content \citep{gioandhaynes85}. The stripping of gas from  this galaxy pair has significantly contributed to the intracluster gas in the region \citep{huangsarazin96}. In the inner 300 arcsec ($\sim$228 kpc) region of A2151C(F), the emission is weaker compared to that in A2151C(B) but the surface brightness seems to be similar in the outer parts ($>$ 300 arcsec) of the two subclumps.

\begin{figure*}
\begin{multicols}{2}
\subcaptionbox{}{\includegraphics[width=\linewidth,height=5.4cm]{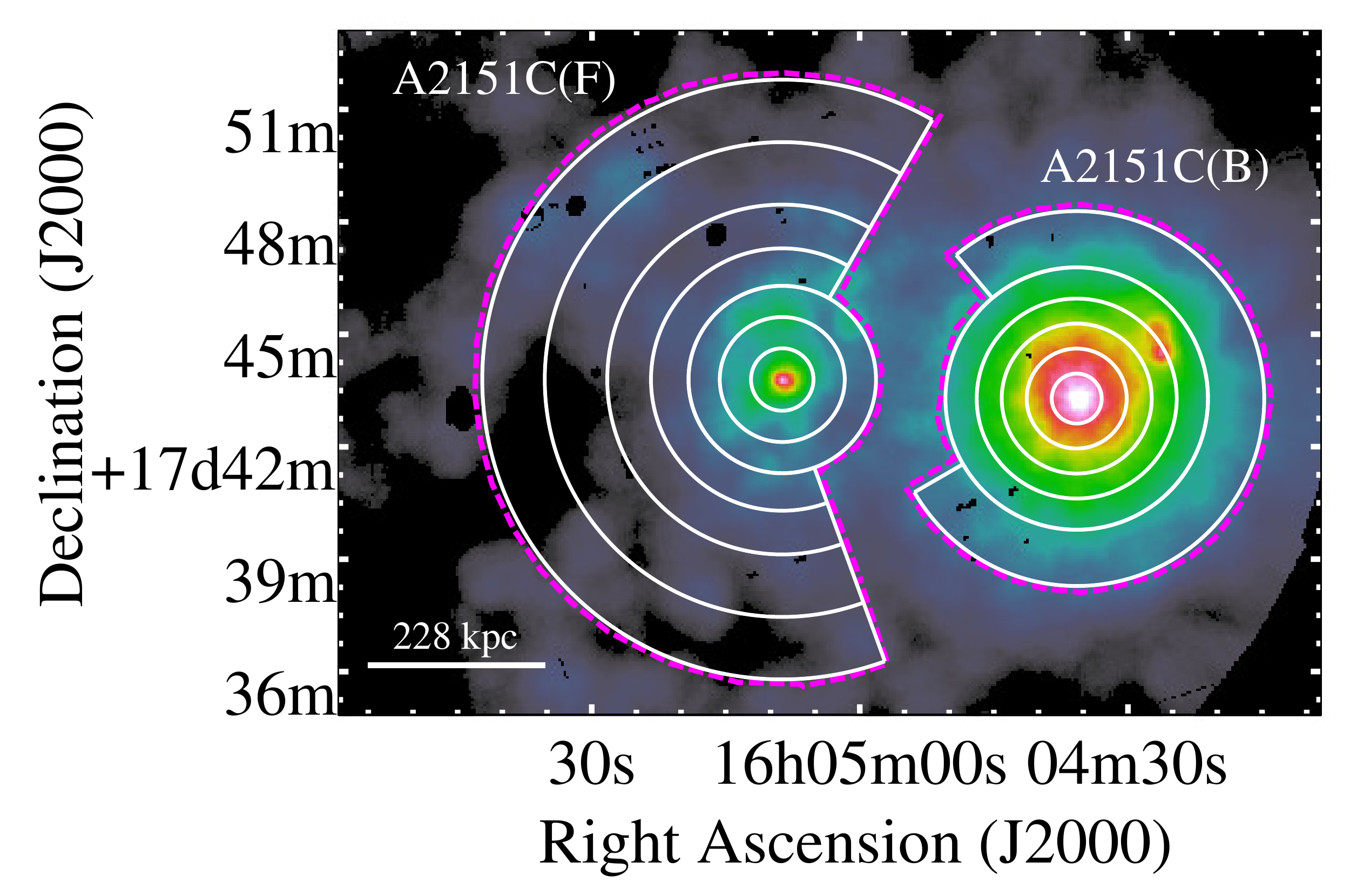}}\par
\subcaptionbox{}{\includegraphics[width=0.7\linewidth,height=8.9cm,angle=-90]{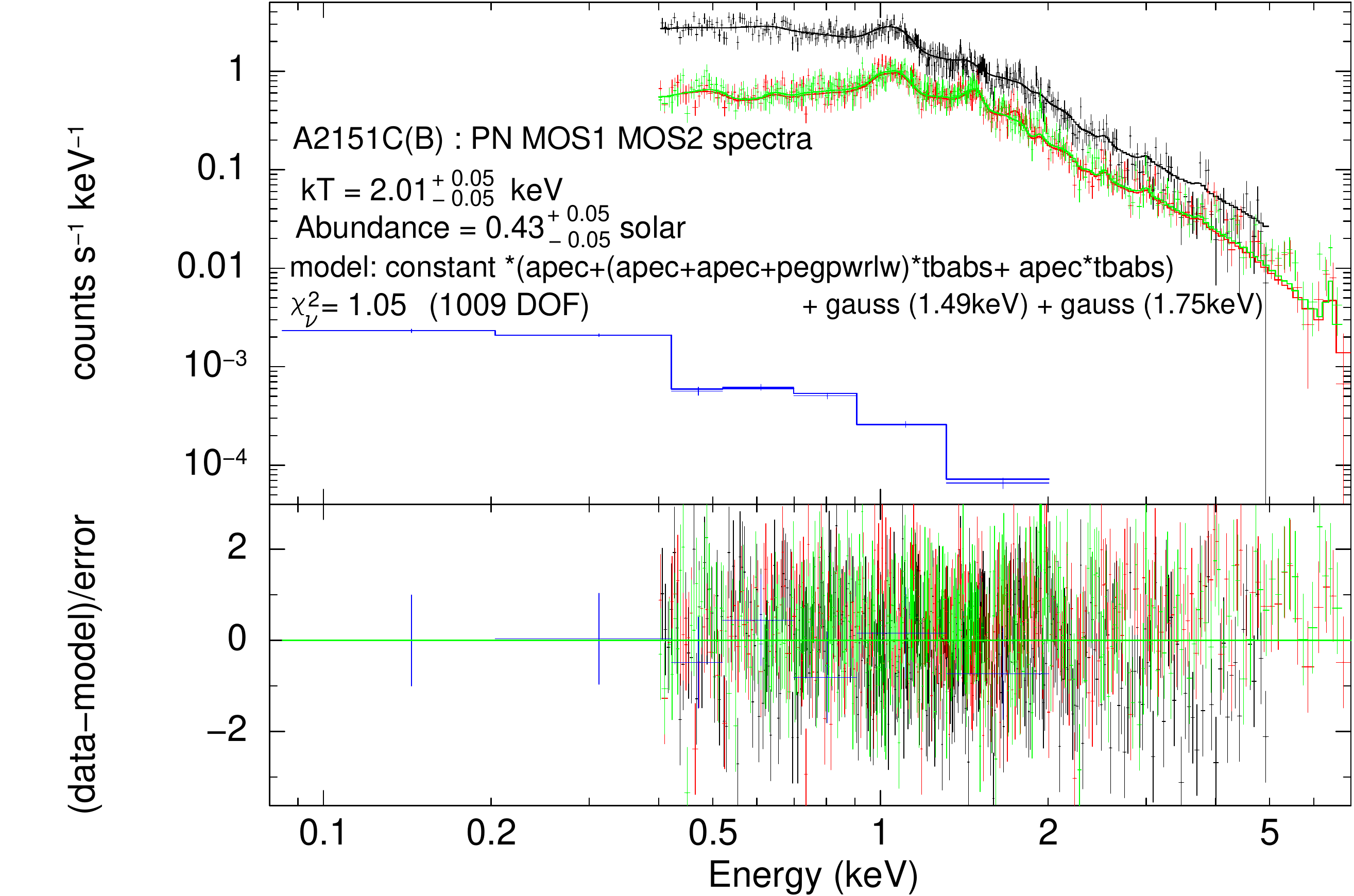}}\par
\end{multicols}
\begin{multicols}{2}
\subcaptionbox{}{\includegraphics[width=0.7\linewidth,height=8.9cm,angle=-90]{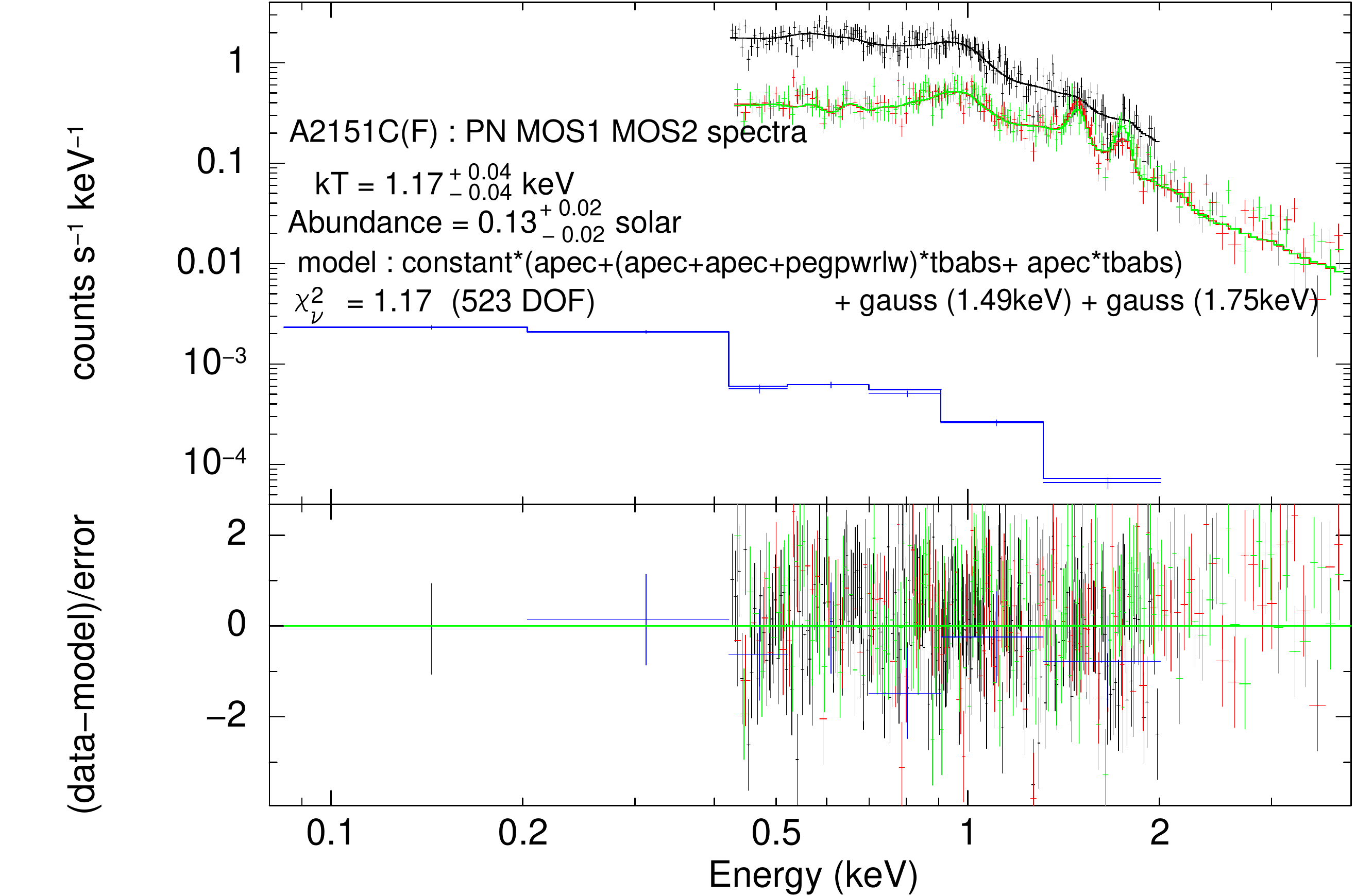}}\par
\subcaptionbox{}{\includegraphics[width=0.7\linewidth,height=8.9cm,angle=-90]{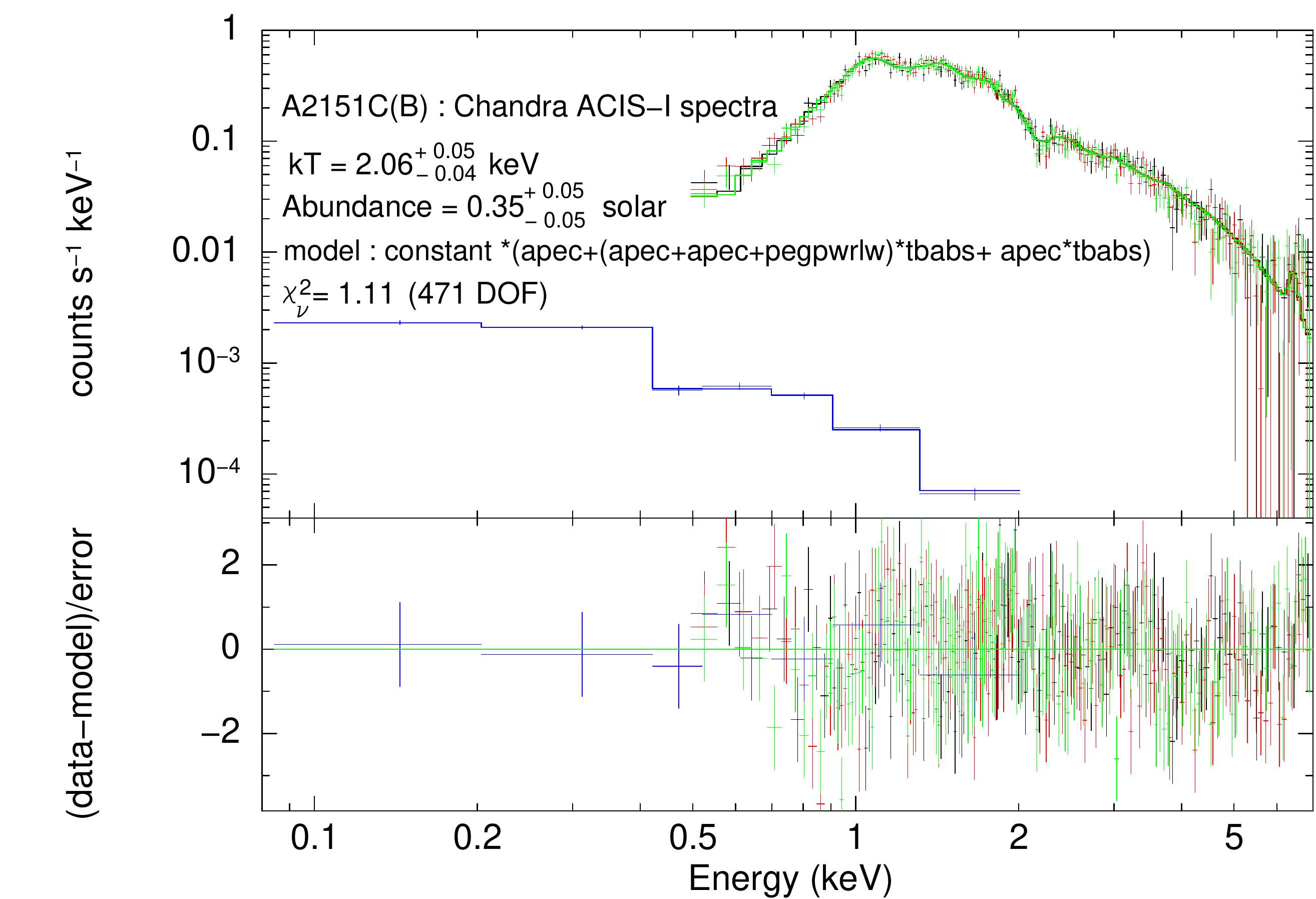}}\par
\end{multicols}
\caption{\textbf{(a)}: Regions used for extraction of spectra are shown. Those outlined in \textit{magenta} color were used for the PN, MOS1, and MOS2 \textit{XMM-Newton} spectral analysis of the full A2151C(B) (\textit{west : circle of radius 210} arcsec \textit{+ a 280\textdegree{} wide sector with inner and outer radii of 210 and 300} arcsec \textit{respectively}) and A2151C(F) (\textit{east : circle of radius 150} arcsec \textit{+ a 230\textdegree{} wide sector with inner and outer radii of 150 and 480} arcsec \textit{respectively}). The same \textit{magenta} region towards the \textit{west} was also used for performing the average spectral analysis of A2151C(B) with \textit{Chandra} data. Regions shown in \textit{white} color were used in the azimuthally averaged spectral analysis of A2151C(B) and A2151C(F) using \textit{XMM-Newton} and \textit{Chandra} data. \textbf{(b)}: Average spectra of A2151C(B) obtained from the \textit{XMM-Newton} PN (black), MOS1 (red), and MOS2 (green) detectors. The spectra were fitted simultaneously with the RASS diffuse background spectrum (shown in blue) using the model \textit{constant*(apec+(apec+apec+pegpwrlw)*tbabs + apec*tbabs)} \textit{+ gauss} (1.49 keV) \textit{+ gauss} (1.75 keV) in the energy range 0.4--5.0 keV for PN and 0.4--7.0 keV for MOS detectors. \textbf{(c)}: Average spectra of A2151C(F) obtained from the \textit{XMM-Newton} PN (black), MOS1 (red), and MOS2 (green) detectors. The spectra were fitted simultaneously with the RASS data (shown in blue) using the model \textit{constant*(apec+(apec+apec+pegpwrlw)*tbabs + apec*tbabs)} \textit{+ gauss} (1.49 keV) \textit{+ gauss} (1.75 keV) model in the energy range 0.4--2.0 keV for PN and 0.4--4.0 keV for MOS detectors. \textbf{(d)}: Average spectra of A2151C(B) obtained using the three \textit{Chandra} ACIS-I observations -- ObsID 19592 (black), ObsID 20086 (red), and ObsID 20087 (green). The spectra were fitted simultaneously with the RASS spectrum (shown in blue) using the model \textit{constant*(apec+(apec+apec+pegpwrlw)*tbabs + apec*tbabs)} in the energy range 0.4--7.0 keV. The lower panels in subplots (b)--(d) show the residuals in terms of sigmas with error bars of size one.}
\label{fig:fig2}
\end{figure*}

\subsection{Average X-ray spectra}\label{sec3.2}
\subsubsection{\textbf{XMM-Newton}}\label{sec3.2.1} The fov of 0\textdegree{}.5 diameter of \textit{XMM-Newton} EPIC  covers both A2151C(B) and A2151C(F) as seen in Fig. \ref{fig:fig1}\textcolor{blue}{(a)}. Average X-ray spectra of the two visible X-ray clumps were extracted from regions outlined in \textit{magenta} color in \autoref{fig:fig2}\textcolor{blue}{(a)}. This was done for each of the three EPIC detectors. The regions were centred on the X-ray surface brightness peaks of the two clumps. It can be seen in Fig. \ref{fig:fig1}\textcolor{blue}{(a)} that the X-ray emission from the two groups is not well separated at least in the 2D projected space. We, therefore, selected regions highlighted in \textit{magenta} color in Fig. \ref{fig:fig2}\textcolor{blue}{(a)} in order to avoid contamination of the average X-ray spectral properties of one group due to the other. The extraction region for A2151C(B) was chosen as a circle of radius 210 arcsec plus a 280$^{\circ}$ wide annular sector of inner and outer radii of 210 and 300 arcsec respectively centred at RA = 16$^{h}$04$^{m}$35$^{s}$.73 and Dec = + 17\textdegree{}43$'$17$''$.31. A region comprising a circle of radius 150 arcsec and a sector of angular width 230$^{\circ}$ with inner and outer radii of 150 and 480 arcsec respectively, centred at RA = 16$^h$05$^m$08$^s$.70 and Dec = +17\textdegree{}43$'$47$''$.96, was selected for the global spectral extraction of A2151C(F). Background spectra and spectral responses in these regions were generated as described in \S2.1.

The PN, MOS1, and MOS2 spectra thus extracted were fitted jointly with the RASS diffuse background spectrum (\S2.1) using a common set of XSPEC models: \textit{constant*(apec+(apec+apec+pegpwrlw)*tbabs + apec*tbabs)} + \textit{gauss} (1.49 keV) + \textit{gauss} (1.75 keV). For consistency with the RASS spectral data which is in units of cts s$^{-1}$ arcmin$^{-2}$, a \textit{constant} factor was used to scale down the model normalisation of the PN, MOS1, and MOS2 data. The \textit{constant} factor equals the active sky area available for each detector in arcmin$^2$ units\footnote{This information can be derived from the BACKSCAL keyword in the spectral file header.}. The model component  \textit{apec+(apec+apec+pegpwrlw)*tbabs} accounts for the complete XRB (i.e., SXRB+CXB). The three \textit{apec} components here model the SXRB contribution -- the unabsorbed thermal emission (kT fixed at 0.1 keV) from the LHB, the absorbed thermal emission (kT fixed at 0.1 keV) from the GH, and the absorbed thermal emission from the NPS (kT fixed at 0.235 keV\footnote{This value was obtained from fitting the RASS diffuse background spectrum with the XRB model.}) (\S2.1). The \textit{pegpwrlw} component accounts for the Cosmic X-ray Background (CXB) which was approximated by a power-law with a photon index of 1.42 and had its normalisation fixed based on the 2--10 keV CXB flux value ($2.15\pm{0.26} \times 10^{-11} \text{erg cm}^{-2} \text{s}^{-1} \text{deg}^{-2}$) obtained by \citet{lumb2002}. 
The component \textit{apec*tbabs} models the absorbed X-ray emission from the cluster. The two \textit{Gaussian} terms are used to model the instrumental background lines -- the Al K$\alpha$ line at $\sim$1.49 keV for PN, MOS1 and MOS2, and the Si K$\alpha$ line at $\sim$1.75 keV for MOS1 and MOS2 detectors.  It is to be noted that we used the \textit{XMM-Newton} modelling of the QPB as mentioned in \S2.1. Therefore, the contribution of the instrumental background lines to the total background was modelled separately. For the RASS background spectrum, the \textit{constant} factor was set equal to 1.0 and the normalisation value of the \textit{apec} model describing the cluster emission and of the gaussian terms was set to 0.0. The model component \textit{apec} is the astrophysical plasma emission model of \citet{apecsmith2001} and \textit{tbabs} is the Tuebingen-Boulder interstellar medium (ISM) absorption model. The \textit{tbabs} model calculates the cross-section for X-ray absorption by the ISM as the sum of the cross-sections for X-ray absorption due to the gas-phase ISM, the grain-phase ISM, and the molecules in the ISM. The elemental abundances used by \textit{tbabs} are as given by \citet{wilms2001}. The total hydrogen column density (neutral+molecular) along the line of sight (LOS) to the cluster, $N_\mathrm{H}$ was taken to be 3.91 x 10$^{20}$ cm$^{-2}$ based on the work of \citet{willingale2013}\footnote{\url{https://www.swift.ac.uk/analysis/nhtot} : The tool returns the neutral, molecular and total galactic column density of Hydrogen.}. The temperature, abundance and  normalisation of the \textit{apec} model describing the cluster emission were allowed to vary freely during the fit, and the redshift value was frozen at 0.0368 (SIMBAD astronomical database : \citet{zabludoff93}). In the case of A2151C(B), we performed a joint fit in the energy range 0.4--7.0 keV for the MOS data and 0.4--5.0 keV for the PN data. The energy range 0.4--4.0 keV was used for A2151C(F). The reason for selecting a smaller energy band for A2151C(F) was that significant X-ray source emission was observed only up to 2.0 keV for the PN spectrum and up to  4.0 keV for MOS1 and MOS2 spectra.  The PN spectrum had significantly higher modelled QPB as compared to the MOS1 and MOS2 spectra.  The data along with the model spectra for A2151C(B) and A2151C(F) are shown in \autoref{fig:fig2}\textcolor{blue}{(b)} and \autoref{fig:fig2}\textcolor{blue}{(c)} respectively. 

\subsubsection{\textbf{Chandra}}\label{sec3.2.2} The region used for the extraction of spectrum for A2151C(B) from  \textit{Chandra} data is the same as that used for the average spectral analysis with \textit{XMM-Newton} data. Spectra and responses were extracted from each of the three \textit{Chandra} ACIS-I observations as explained in \S2.2. It was not possible to perform the spectral analysis of A2151C(F) (as done with the \textit{XMM-Newton} data) since the \textit{Chandra} observations covered it only partially due to the smaller fov of ACIS-I. The  spectra in the energy range 0.4--7.0 keV obtained from the three \textit{Chandra} ACIS-I observations were fitted simultaneously with the RASS diffuse background spectrum using the XSPEC model: \textit{constant*(apec+(apec+apec+pegpwrlw)*tbabs + apec*tbabs)}. The individual model component definitions, and the $N_\mathrm{H}$ and redshift values are the same as those used in \S3.2.1. The photon index of the \textit{pegpwrlw} component was set to 1.4 and its normalisation was fixed based on the 2--8 keV CXB flux value ($1.7\pm{0.2} \times 10^{-11} \text{erg cm}^{-2} \text{s}^{-1}\text{deg}^{-2}$) obtained by \citet{hickox2006}. Although blanksky observations were used to generate background spectra, these spectra were scaled based on the ratio of observation to total background counts in 9.0-12.0 keV energy range corresponding primarily to the particle background.  Hence, the \textit{pegpwrlw} model was required to separately account for the CXB emission contribution. The data along with the model spectra are shown in \autoref{fig:fig2}\textcolor{blue}{(d)}. 

The best-fitting values of gas temperature, abundance, and \textit{apec} normalisation are provided in \autoref{tab:tab2}. The temperature and abundance values obtained for A2151C(B) with the analysis done using \textit{XMM-Newton} and \textit{Chandra} data agree quite well with each other.
\begin{table*}
 \begin{center}
  \captionsetup{justification=centering}
  \caption{Best-fitting parameters -- temperature (kT), abundance (Z) and apec normalisation ($\mathcal{N}$) -- obtained from the X-ray spectral analysis of the full A2151C(B) and A2151C(F) regions (outlined in \textit{magenta} in Fig. \ref{fig:fig2}\textcolor{blue}{(a)}). The minimum reduced $\chi^2$ statistic along with the degrees of freedom (DOF) is also listed. The \textit{XMM-Newton} and \textit{Chandra} global spectra were fitted using the models \textit{constant*(apec+(apec+apec+pegpwrlw)*tbabs + apec*tbabs) + gauss}(1.49 keV)\textit{ + gauss}(1.75 keV) and \textit{constant*(apec+(apec+apec+pegpwrlw)*tbabs + apec*tbabs)} respectively, simultaneously with the RASS diffuse background spectrum. The error bars correspond to 90 per cent confidence intervals.}
  \begin{tabular}{ccccccc}
    \hline
\hline
Region & Data & kT & Z & apec norm. ($\mathcal{N}$) & X-ray Luminosity ($L_X$) & $(\chi_{\nu}^2)_{\text{min}}${\hskip 0.04in}(DOF)\\
&&(keV)&(Z$_\odot$)& ($10^{-2}$ cm$^{-5}$)& ($10^{43}$ erg s$^{-1}$)&\\
\\
\hline
\hline
A2151C(B)&\textit{XMM-Newton}&$2.01\pm{0.05}$&$0.43\pm{0.05}$&$1.02\pm{0.02}$&$3.03^{+0.02}_{-0.04}$ \hspace{0.38cm}(0.4$-$7.0 keV) &1.05 (1009)\\
A2151C(F)&\textit{XMM-Newton}&$1.17\pm{0.04}$&$0.13\pm{0.02}$&$0.43\pm{0.02}$&$1.13\pm{0.02}$ \hspace{0.38cm}(0.4$-$7.0 keV) &1.17 (523)\\
\hline
A2151C(B)&\textit{Chandra}&$2.06^{+0.05}_{-0.04}$&$0.35\pm{0.05}$&$1.08\pm{0.03}$&$3.23^{+0.02}_{-0.03}$ \hspace{0.38cm}(0.4$-$7.0 keV) &1.11 (471)\\
\hline\hline
  \end{tabular}
   \label{tab:tab2}
  \end{center}
 \end{table*}
 
 \begin{table*}
 \begin{center}
  \captionsetup{justification=centering}
  \caption{Best-fitting parameters obtained from the projected spectral analysis of annuli/sectors in A2151C(B) (0.4--7.0 keV) and A2151C(F) (0.4--4.0 keV) (shown in \textit{white} in Fig. \ref{fig:fig2}\textcolor{blue}{(a)}) using \textit{XMM-Newton} and \textit{Chandra} data. The spectra of all the regions were fitted using the models \textit{constant * (apec+(apec+apec+pegpwrlw)*tbabs + apec*tbabs) +  gauss} (1.49 keV)\textit{ + gauss} (1.75 keV) and \textit{constant * (apec+(apec+apec+pegpwrlw)*tbabs + apec*tbabs)} for \textit{XMM-Newton} and \textit{Chandra} data respectively. The \textit{XMM-Newton} and \textit{Chandra} spectra were simultaneously analysed with the RASS diffuse background spectrum. The regions used for spectral extraction and the temperature (kT), abundance (Z), and derived electron density (n$_{e}$) values are listed. The errors are quoted at 90 per cent confidence level.}
  \begin{tabular}{cccccc}
    \hline
\hline
X-ray group &Region&Data & kT & Z & n$_{e}$\\
&&&(keV)&(Z$_\odot$)&($10^{-3}$cm$^{-3}$)\\
\\
\hline
\hline
A2151C(B)&0--40 arcsec&\textit{XMM-Newton}&$1.83^{+0.08}_{-0.07}$&$0.88_{-0.16}^{+0.19}$&$11.81_{-0.41}^{+0.39}$\\
& &\textit{Chandra}&$2.07\pm{0.07}$&$0.92_{-0.16}^{+0.19}$&$13.12_{-0.44}^{+0.43}$\\
\\
&40--80 arcsec&\textit{XMM-Newton}&$2.30_{-0.13}^{+0.12}$&$0.77_{-0.15}^{+0.18}$&$5.20\pm{0.14}$\\
& &\textit{Chandra}&$2.42\pm{0.10}$&$0.67_{-0.14}^{+0.16}$&$5.35_{-0.15}^{+0.16}$\\
\\
&80--120 arcsec&\textit{XMM-Newton}&$2.28\pm{0.13}$&$0.51_{-0.12}^{+0.14}$&$3.17_{-0.09}^{+0.08}$\\
& &\textit{Chandra}&$2.33_{-0.12}^{+0.13}$&$0.33_{-0.11}^{+0.13}$&$3.28\pm{0.10}$\\
\\
&120--160 arcsec&\textit{XMM-Newton}&$2.36_{-0.16}^{+0.17}$&$0.27_{-0.13}^{+0.12}$&$2.26_{-0.06}^{+0.07}$\\
& &\textit{Chandra}&$2.16_{-0.13}^{+0.14}$&$0.22_{-0.10}^{+0.12}$&$2.24\pm{0.08}$\\
\\
&160--210 arcsec&\textit{XMM-Newton}&$2.06\pm{0.15}$&$0.20_{-0.08}^{+0.09}$&$1.56\pm{0.04}$\\
& &\textit{Chandra}&$1.92_{-0.13}^{+0.14}$&$0.19_{-0.09}^{+0.11}$&$1.57\pm{0.06}$\\
\\
&210--300 arcsec (280\textdegree{} wide sector)&\textit{XMM-Newton}&$1.90_{-0.15}^{+0.16}$&$0.14_{-0.07}^{+0.09}$&$0.94\pm{0.03}$\\
& &\textit{Chandra}&-&-&-\\
\hline
A2151C(F)&0--50 arcsec&\textit{XMM-Newton}&$1.07\pm{0.04}$&$0.26_{-0.05}^{+0.07}$&$4.67_{-0.27}^{+0.25}$\\
\\
&50--100 arcsec&&$1.20_{-0.06}^{+0.07}$&$0.22_{-0.06}^{+0.07}$&$1.93\pm{0.10}$\\
\\
&100--150 arcsec&&$1.32\pm{0.10}$&$0.13_{-0.04}^{+0.05}$&$1.42\pm{0.06}$\\
\\
&150--210 arcsec (230\textdegree{} wide sector)&&$1.26\pm{0.10}$&$0.16_{-0.06}^{+0.08}$&$0.96\pm{0.05}$\\
\\
&210--280 arcsec (230\textdegree{} wide sector) &&$1.34_{-0.13}^{+0.14}$&$0.17_{-0.08}^{+0.11}$&$0.67\pm{0.04}$\\
\\
&280--380 arcsec (230\textdegree{} wide sector) &&$1.15_{-0.12}^{+0.14}$&$0.02_{-0.00}^{+0.03}$&$0.59\pm{0.04}$\\
\\
&380--480 arcsec (230\textdegree{} wide sector) &&$1.04_{-0.11}^{+0.13}$&$0.02\pm{0.02}$&$0.48\pm{0.03}$\\
\hline\hline
  \end{tabular}
   \label{tab:tab3}
  \end{center}
  \end{table*}
 
  \begin{figure*}
\begin{multicols}{3}
\subcaptionbox{}{\includegraphics[width=\linewidth,height=4cm]{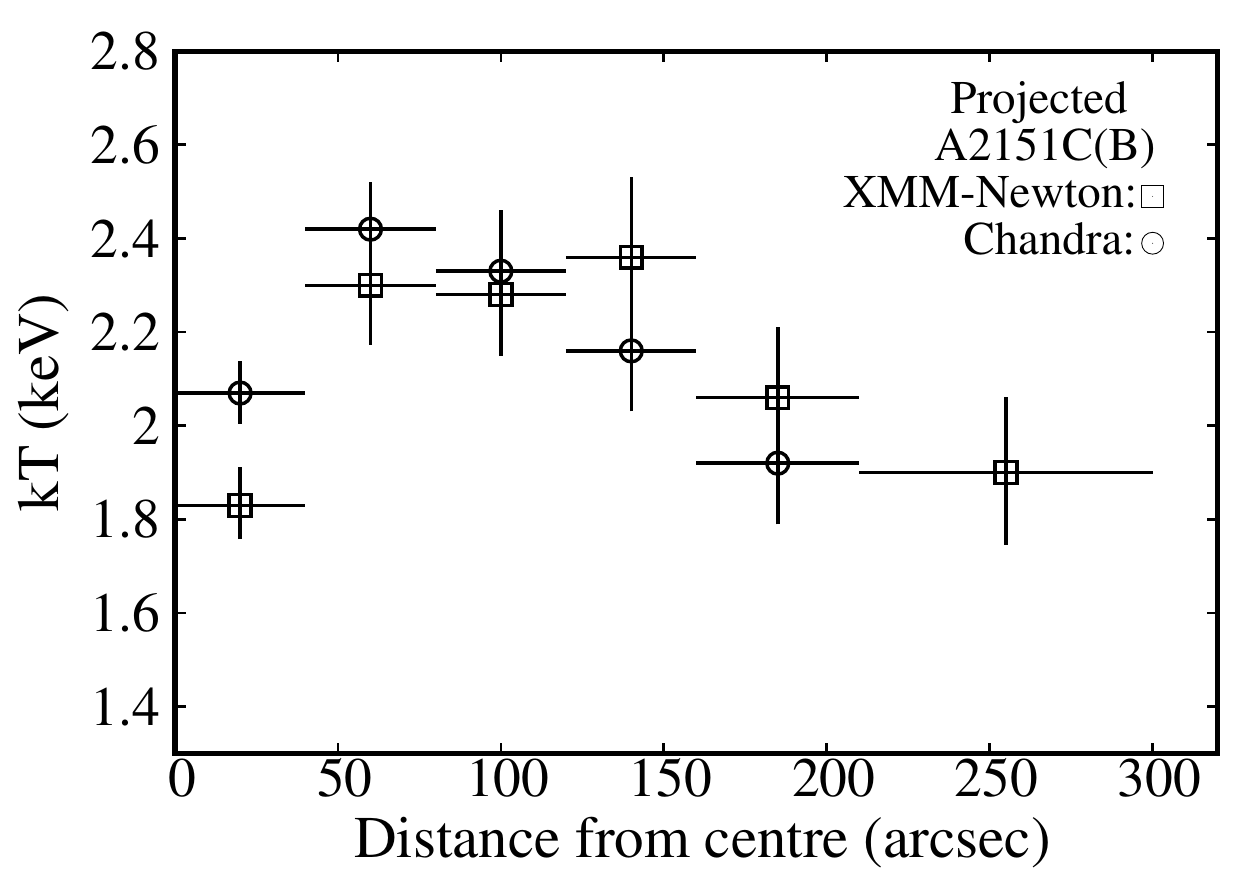}}\par
\subcaptionbox{}{\includegraphics[width=\linewidth,height=4cm]{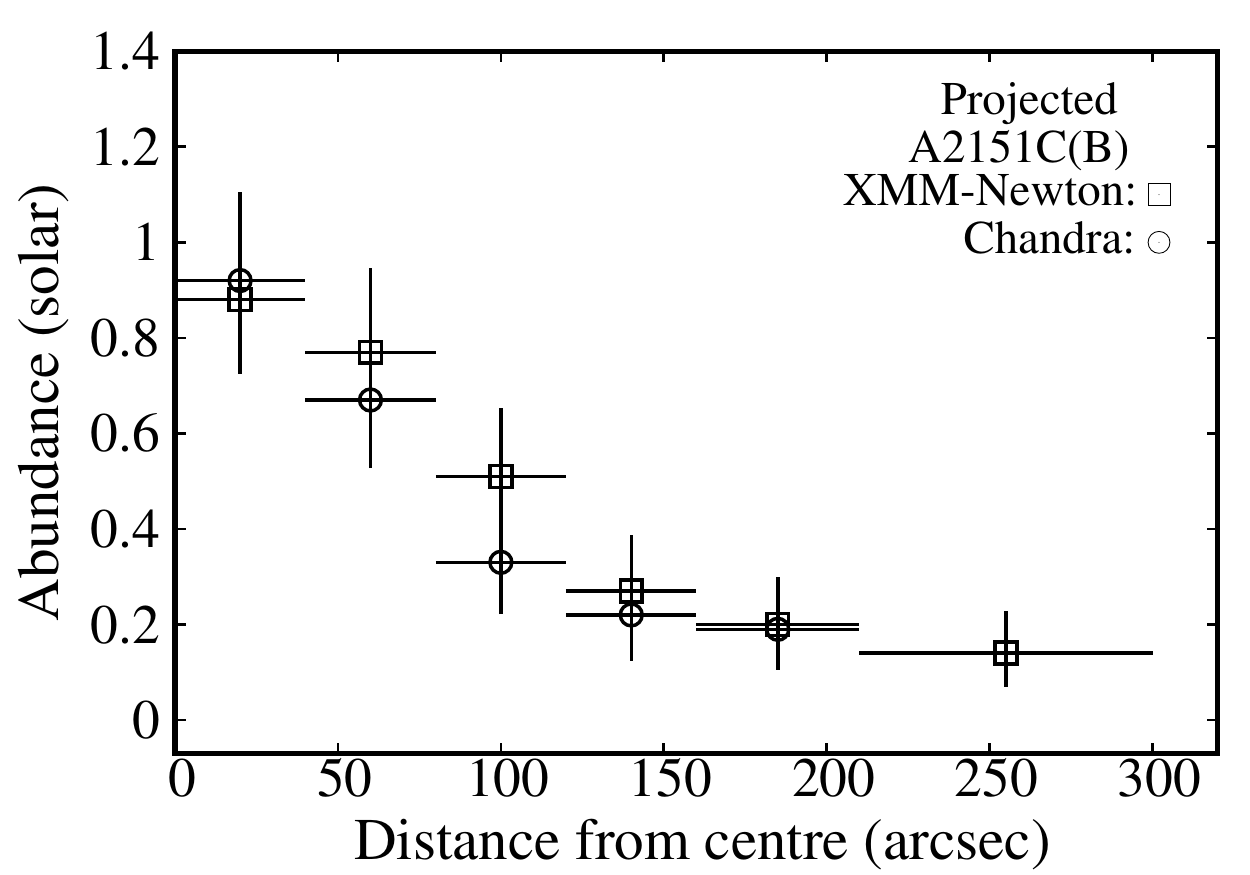}}\par
\subcaptionbox{}{\includegraphics[width=\linewidth,height=4cm]{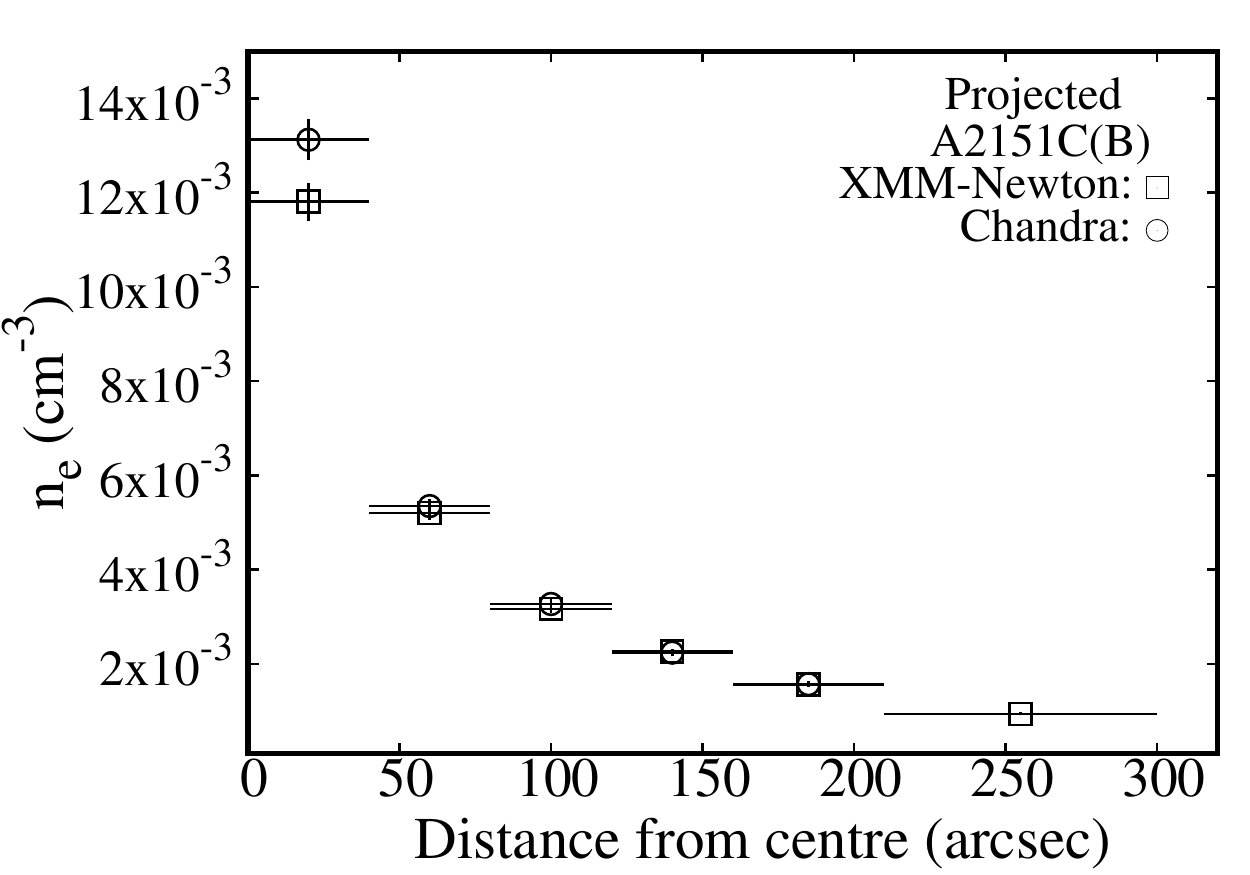}}\par
\end{multicols}
\begin{multicols}{3}
\subcaptionbox{}{\includegraphics[width=\linewidth,height=4cm]{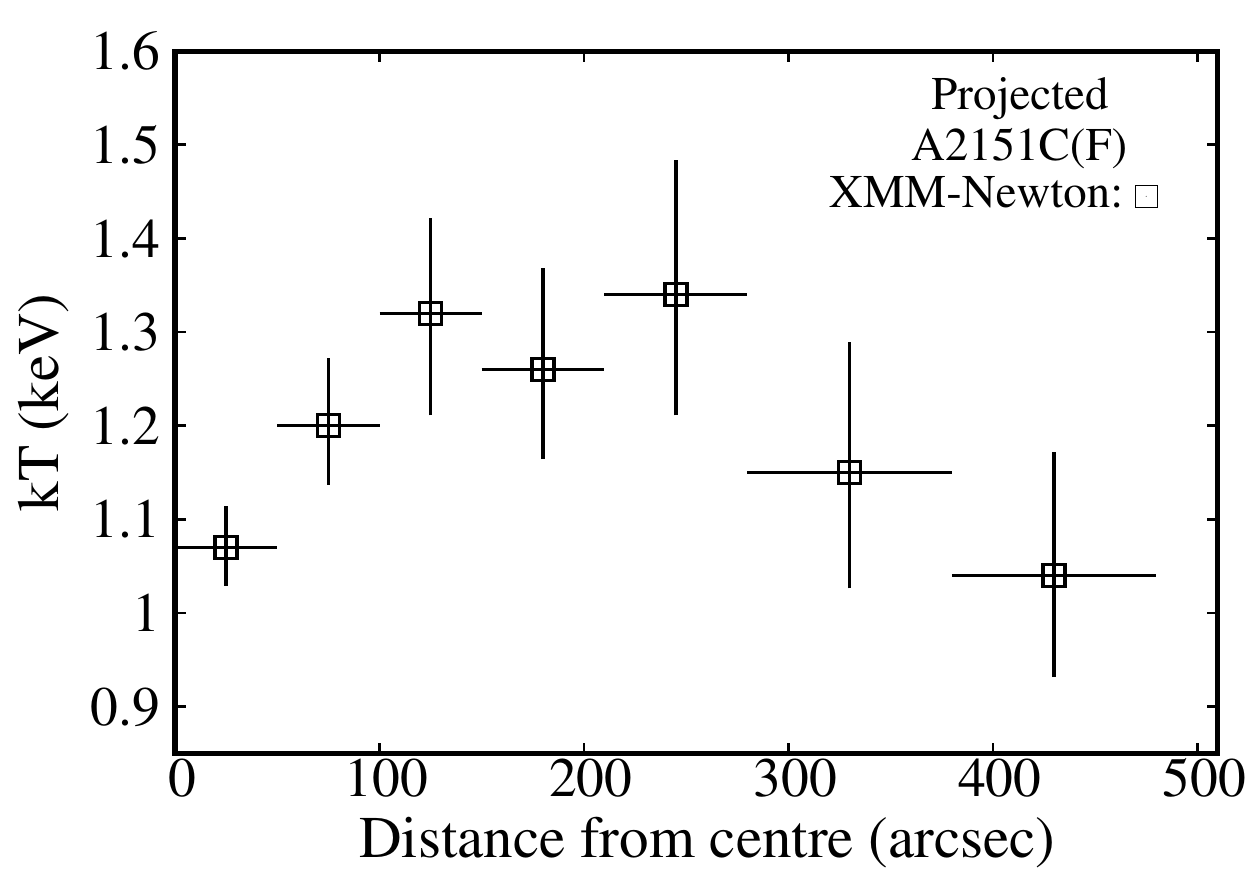}}\par
\subcaptionbox{}{\includegraphics[width=\linewidth,height=4cm]{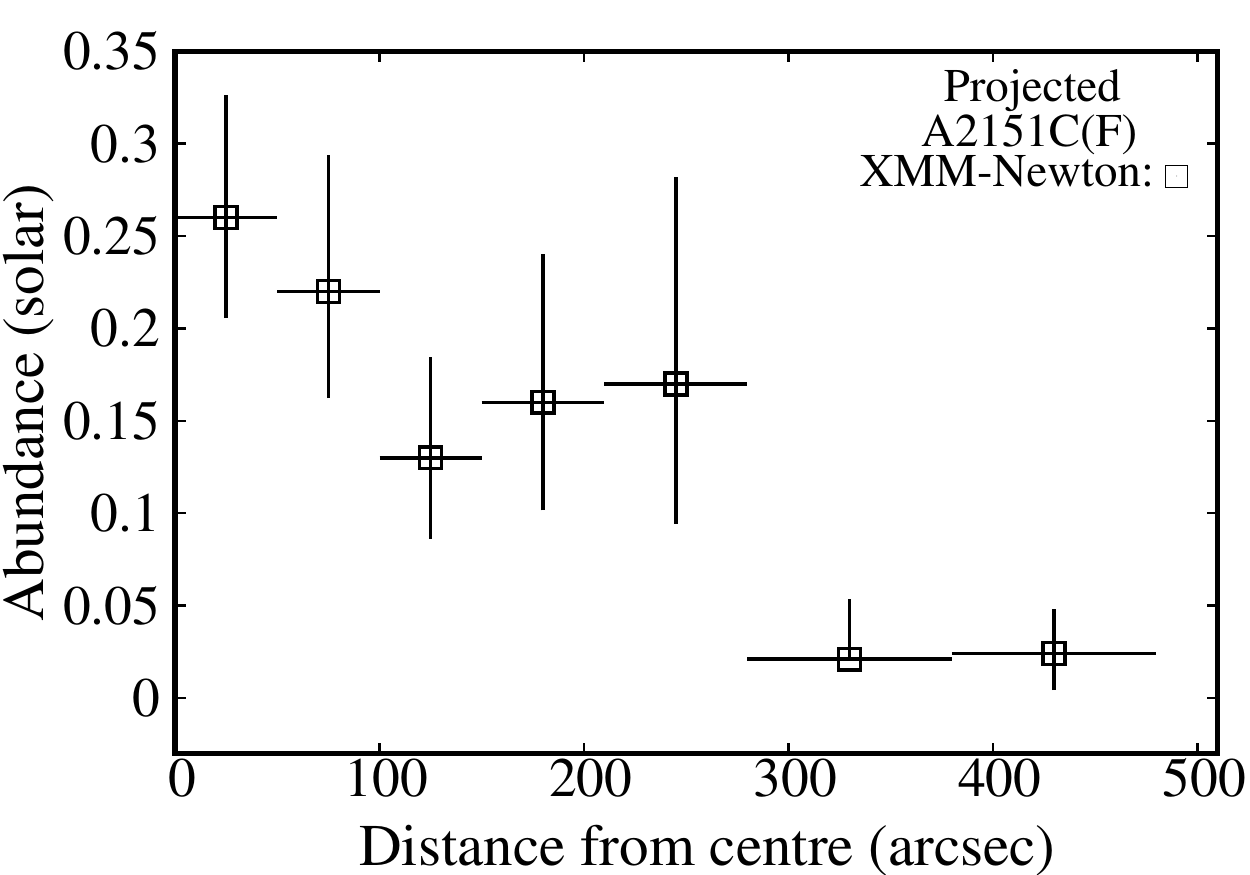}}\par
\subcaptionbox{}{\includegraphics[width=\linewidth,height=4cm]{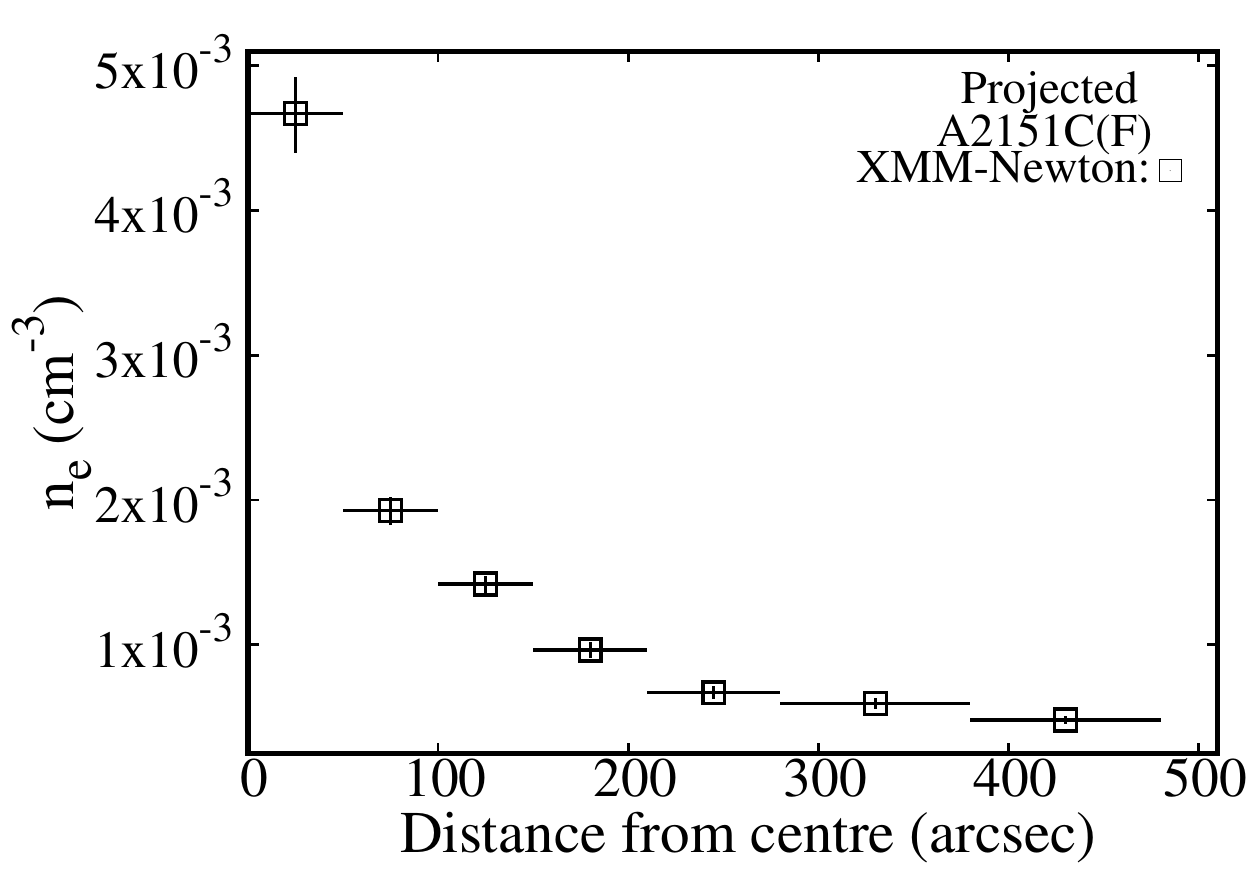}}\par
\end{multicols}
\caption{Projected temperature (kT), abundance, and electron density (n$_{e}$) profiles of A2151C(B) and A2151C(F) obtained from the analysis of spectra extracted in different annuli/sectors (shown in \textit{white} in Fig. \ref{fig:fig2}\textcolor{blue}{(a)}) using \textit{XMM-Newton} (open box) and \textit{Chandra} (open circle) data. The error bars correspond to a 90 per cent confidence interval.}
\label{fig:fig3}
\end{figure*}

\subsection{Azimuthally averaged spectral analysis: Determination of radial profiles of gas temperature, abundance and density}
\subsubsection{\textbf{2D Projected Profiles}}\label{sec3.3.1}Radial profiles of gas temperature and abundance were obtained for A2151C(B) and A2151C(F) by extracting spectra in regions shown in \textit{white} color in Fig. \ref{fig:fig2}\textcolor{blue}{(a)}. The regions were centred at the peak of X-ray emission for both A2151C(B) and A2151C(F). A2151C(B) was divided into six annular regions with outer radii 40, 80, 120, 160, 210, and 300 arcsec for analysis with \textit{XMM-Newton} data. The outermost region (inner and outer radii of 210 and 300 arcsec respectively) was chosen as a sector (and not a full annulus) with an angular width of 280\textdegree{} to avoid mixing of projected gas properties of the two X-ray groups. The region beyond 300 arcsec was not covered fully by the PN and MOS2 chips. A joint spectral analysis with PN, MOS1 and MOS2 data was therefore not possible beyond 300 arcsec. Analysis with \textit{Chandra} data was done using the same regions except the outermost sector which was not included due to poor statistics. A2151C(F) was divided into seven regions with outer radii 50, 100, 150, 210, 280, 380, and 480 arcsec. The four outermost regions with outer radii 210, 280, 380, and 480 arcsec were restricted to sectors with angular width 230\textdegree{}. Since the \textit{Chandra} observations did not cover A2151C(F) fully, only \textit{XMM-Newton} data was used in its analysis.

The spectral modelling followed the same procedure as described in \S3.2.1. The results of spectral fitting are provided in \autoref{tab:tab3}. 
Temperature  and metallicity profiles were obtained directly from the spectral analysis. To derive the electron density (n$_{e}$), we used the cluster \textit{apec} normalisation, $\mathcal{N}$ defined as \begin{equation}\mathcal{N}=\frac{10^{-14}}{4\pi[D_{A}(1+z)]^{2}}\int n_{e}n_{H} dV \hspace{0.6cm}  (cm^{-5})\label{eq:eq1}\end{equation} where $D_{A}$ is the angular diameter distance to the source (cm), $z$ is the redshift, $V$ is the volume of the region, and $n_{e}$ and $n_{H}$ are the electron and hydrogen densities (cm$^{-3}$), respectively.

The XSPEC \textit{apec} model takes into account the contribution from the following elements -- C, N, O, Ne, Mg, Al, Si, S, Ar, Ca, Fe, Ni, and fixes the H and He contributions to the cosmic value, while determining the metal abundance of the plasma. The trace element abundances (Li, Be, B, F, Na, P, Cl, K, Sc,
Ti, V, Cr, Mn, Co, Cu, Zn) are set to the solar value. Including contribution from the elements mentioned above, their relative abundances being determined by the abundance table of \citet{wilms2001}, and assuming a fully ionized plasma, such that each atom of an element contributes Z$_e$ (atomic number of the element) electrons, we derived the $n_H/n_e$ ratio for an ICM abundance of 0.43 Z$_\odot$ (average abundance of A2151C(B) from \textit{XMM-Newton}). This resulted in $n_H/n_e = 0.834$. The $n_H/n_e$ values obtained for ICM abundances of 0.35 Z$_\odot$ (average abundance of A2151C(B) from \textit{Chandra}) and 0.13 Z$_\odot$ (average abundance of A2151C(F)) were similar -- 0.834 and 0.835 respectively. The derived value of the mean molecular weight ($\mu$) was 0.613. We assumed a constant density within each spherical shell and used $n_{H}=0.834n_{e}$ to derive the 2D projected profiles of temperature, abundance and electron density  shown in \autoref{fig:fig3}. It can be seen in Fig. \ref{fig:fig3}\textcolor{blue}{(a)} that the temperature in the central 40 arcsec in A2151C(B) is lower than its surroundings by $\sim$0.3 keV. This confirms that A2151C(B) has a cooling core. In contrast, the temperature drop seen towards the inner regions of A2151C(F) is not significant (Fig. \ref{fig:fig3}\textcolor{blue}{(d)}). There is a slight difference between the temperature values obtained with \textit{XMM-Newton} and \textit{Chandra} in the central 40 arcsec region of A2151C(B). This is possibly the effect of the large chip gaps of the PN and MOS1 detectors, falling within the 0$-$40 arcsec region in the \textit{XMM-Newton} data, in comparison with the \textit{Chandra} ACIS-I detector where no such gaps are seen to coincide with this region.  Fig. \ref{fig:fig3}\textcolor{blue}{(b)} shows that A2151C(B) is fairly metal rich in the inner regions due to chemical enrichment from the BCG NGC 6041. The abundance steadily declines as one goes to the outer parts of the subclump. The abundance of A2151C(F) is much lower than that of A2151C(B). This is an indication of its relatively unevolved nature. Fig. \ref{fig:fig3}\textcolor{blue}{(e)} shows an abrupt rise in the abundance value of A2151C(F) at radii 150--280 arcsec from the centre. Part of this annular region overlaps with the CGG SDSSCGB4240 (Fig. \ref{fig:fig1}\textcolor{blue}{(b)}). The rise in abundance could possibly be due to this galaxy group merging with A2151C(F) (also see Fig. \ref{fig:fig7}\textcolor{blue}{(b)} and \ref{fig:fig7}\textcolor{blue}{(d)}).

\subsubsection{\textbf{Deprojected Profiles}}\label{sec3.3.2}
\begin{figure*}
\begin{multicols}{2}
\subcaptionbox{}{\includegraphics[width=0.85\linewidth,height=5.5cm]{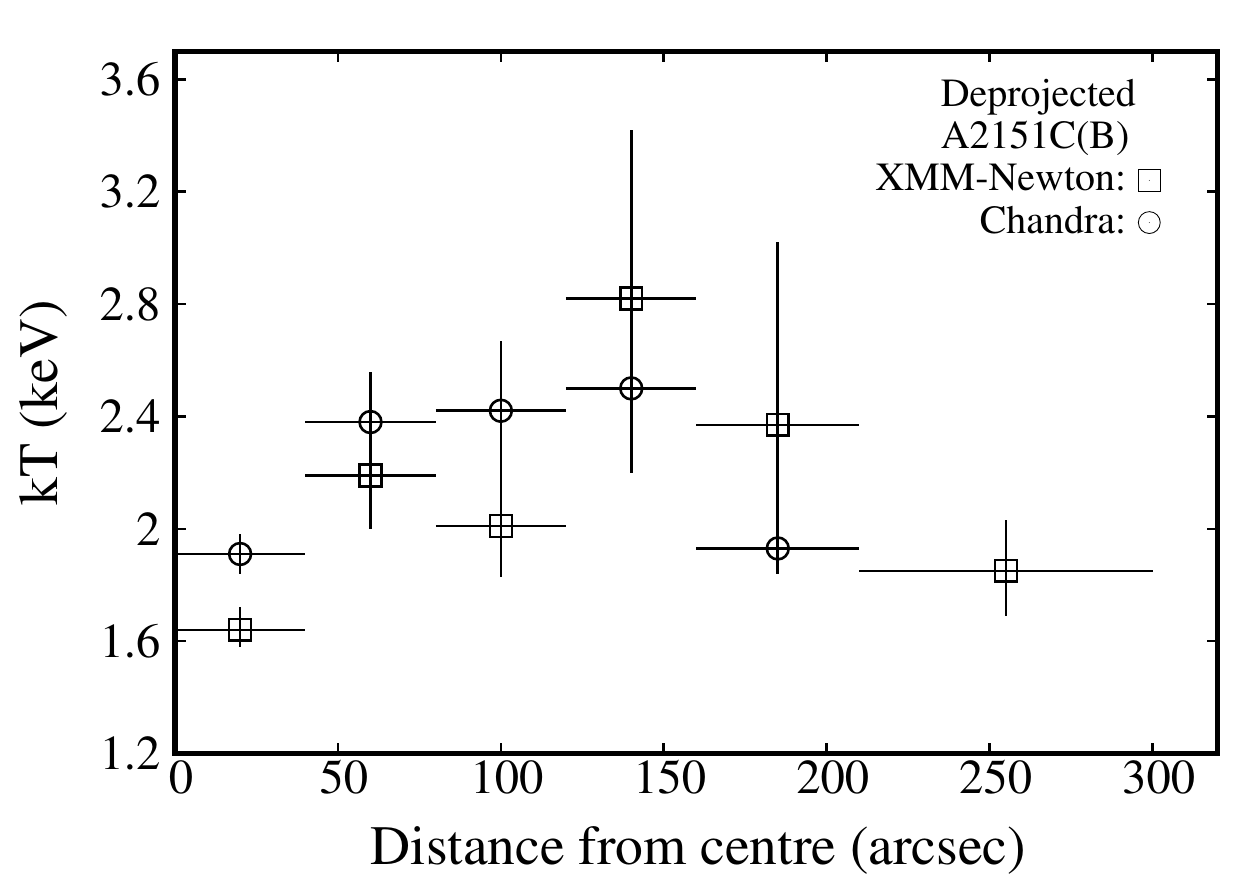}}\par
\subcaptionbox{}{\includegraphics[width=0.85\linewidth,height=5.5cm]{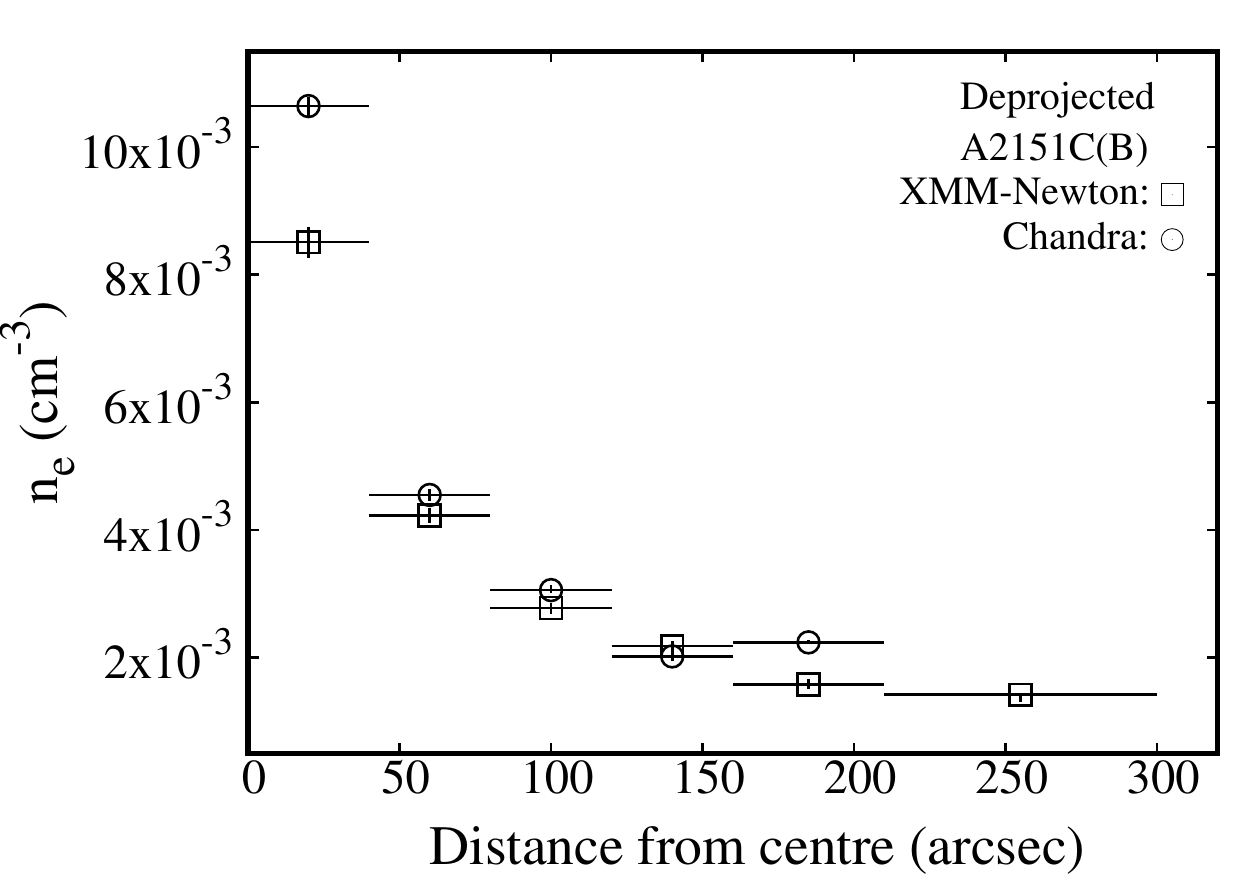}}\par
\end{multicols}
\begin{multicols}{2}
\subcaptionbox{}{\includegraphics[width=0.85\linewidth,height=5.5cm]{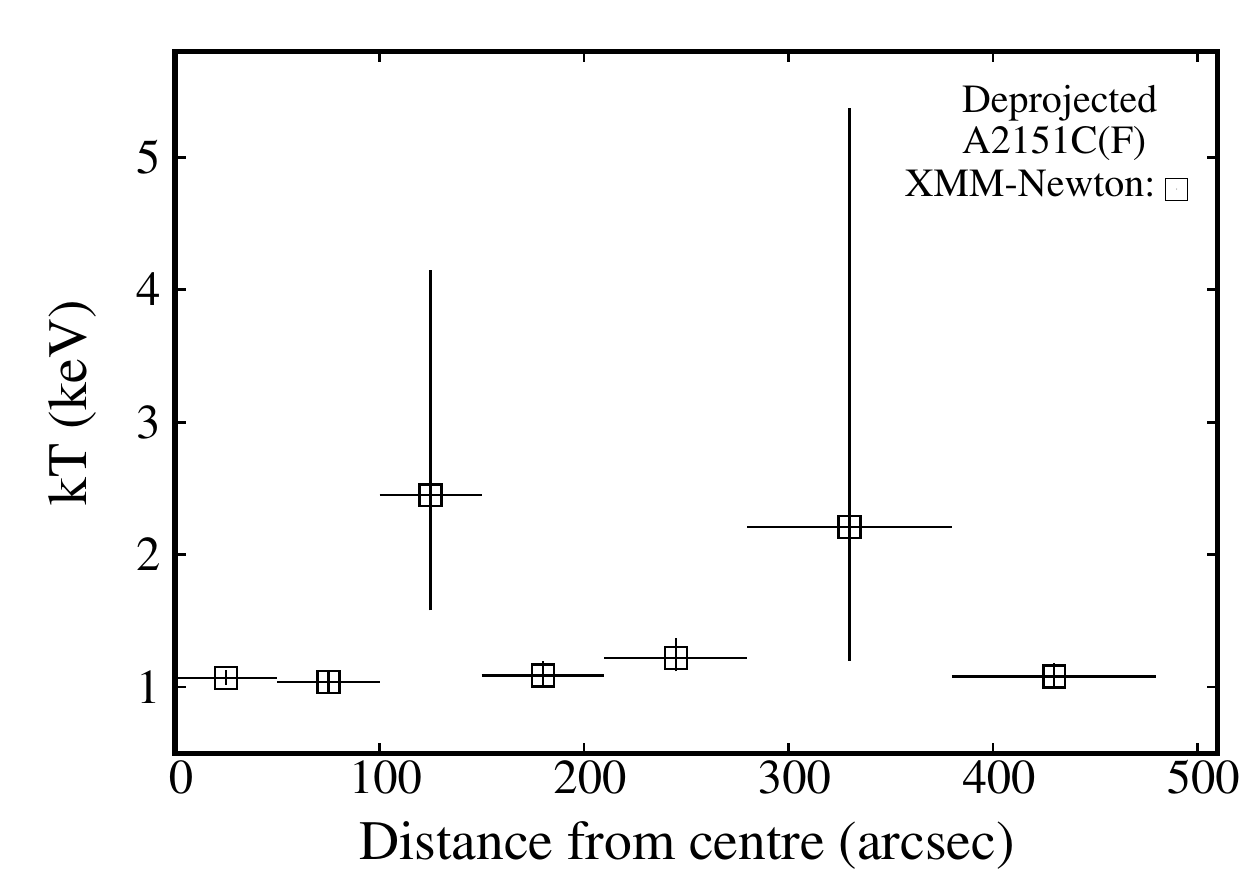}}\par
\subcaptionbox{}{\includegraphics[width=0.85\linewidth,height=5.5cm]{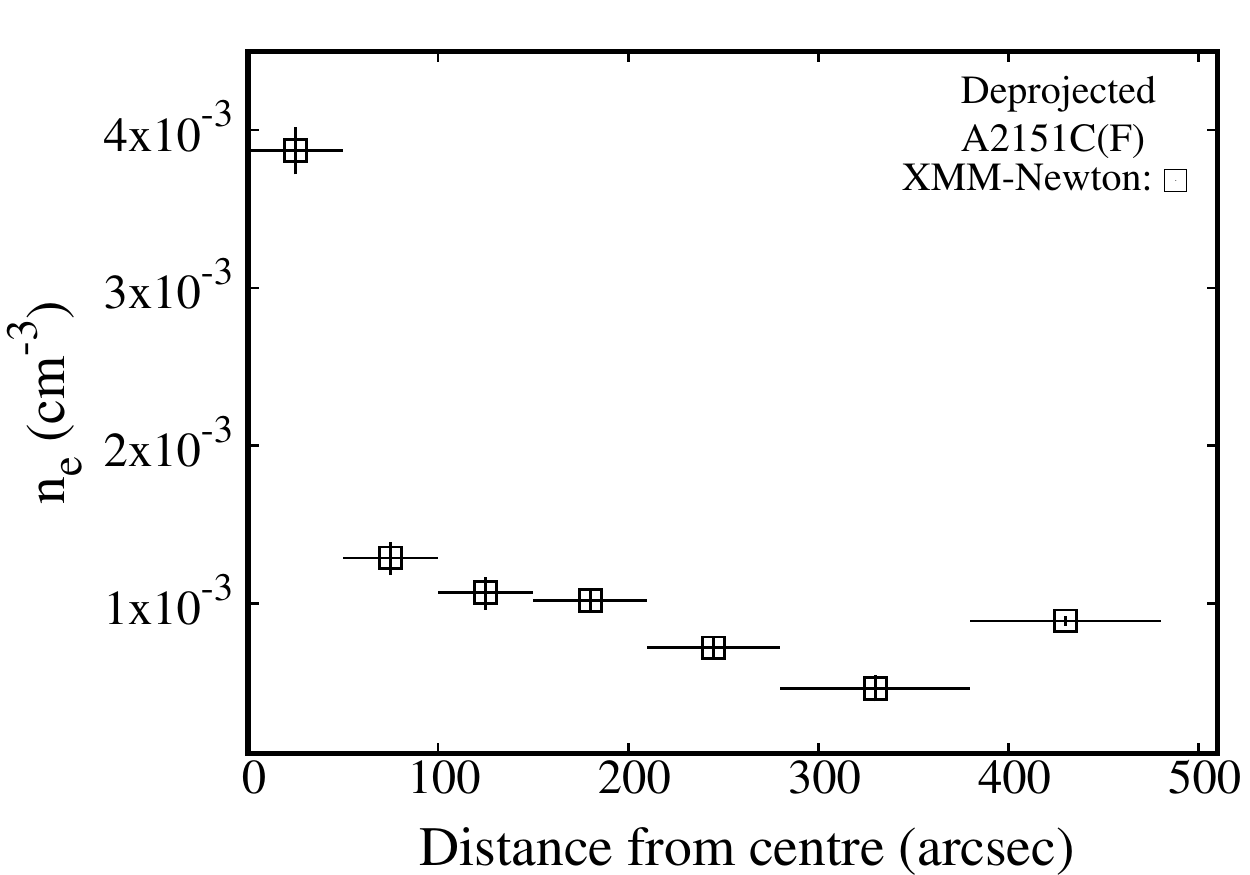}}\par
\end{multicols}
\caption{Deprojected temperature (kT) and electron density (n$_{e}$) profiles of A2151C(B) and A2151C(F) obtained from the spectral analysis with \textit{XMM-Newton} (open box) and \textit{Chandra} (open circle) data. The abundance value in each annulus/sector was kept frozen to the corresponding projected value. The error bars correspond to a 90 per cent confidence interval.}
\label{fig:fig4}
\end{figure*}

When performing a spectral analysis in projection, the inner annuli have significant contribution from the outer gas shells. The variation in the thermodynamic quantities may thus get smoothed out due to these projection effects. We, therefore, carried out a deprojection analysis on the regions described in \S3.3.1. The XSPEC model \textit{projct} which estimates parameters in 3D space from the 2D projected annular spectra, was used for this purpose. We note that the model requires all the spectra belonging to the same annulus/sector to be part of the same data group. The model \textit{apec+(apec+apec+pegpwrlw)*tbabs + projct*(tbabs*apec)} was used for fitting the data. The cluster abundance and XRB normalisation values in each annulus were kept frozen to the corresponding projected values. The deprojection analysis with \textit{XMM-Newton} data also required two additional \textit{gauss} models to account for the instrumental lines for each data group (see \S3.2.1 for model component definitions). The temperature profiles were obtained directly from the spectral analysis. The electron density values were calculated in the same way as described in \S3.3.1. The resulting values are listed in \autoref{tab:tab4} and the profiles are shown in \autoref{fig:fig4}. The errors in the best-fitting parameters are rather large in the deprojected case. A low temperature in the core of A2151C(B) is, however, still seen.

Our analysis uses partial annuli in the outer regions of the two subclusters for reasons mentioned in \S3.2.1. We note that this does not affect the results of the deprojection analysis since the \textit{projct} model uses a normalised volume (defined as the fraction of the ellipsoidal volume intersected by the elliptical annular cross-section) in its calculations. Therefore, if the volume of the ellipsoid is reduced by some amount (resulting from the use of partial annuli) then the physical volume of intersection also gets reduced by the same factor, thus leaving the normalised volume unchanged. In order to account for the use of a combination of complete annuli in the central regions and incomplete annuli towards the outer regions of A2151C(B) and A2151C(F) in the deprojection analysis, we added keywords specifying the start and end angles for the partial annuli to their spectral file extensions. The start and end angles for the outermost partial annulus of A2151C(B) were entered as 210\textdegree{} and 490\textdegree{} respectively while these values were entered as 60\textdegree{} and 290\textdegree{} respectively for the outer four partial annuli of A2151C(F). The start and end angles were measured anticlockwise relative to the x-axis. The following caveats on the use of $projct$ model must be kept in mind, however:\begin{itemize} \item{The model does not account for the effect of missing regions from chip gaps and masked sources.}\item{The outermost bin always has contribution from all of the emission beyond it. The model effectively puts all the extra emission into the outermost bin. This may result in a higher density value in the outermost shell than in the inner regions. The effect cannot be corrected for since this would require a model for the variation of the emission measure and temperature with radius beyond the outermost annulus. In practice, the temperature and density values obtained for the outermost shell with the deprojection analysis are considered as not useful.} \end{itemize}

 \begin{table*}
 \begin{center}
  \captionsetup{justification=centering}
  \caption{Best-fitting parameters obtained from the deprojected spectral analysis of annuli/sectors in A2151C(B) (0.4--7.0 keV) and A2151C(F) (0.4--4.0 keV) (shown in \textit{white} Fig. \ref{fig:fig2}\textcolor{blue}{(a)}). The spectra were fitted using the models \textit{apec+(apec+apec+pegpwrlw)*tbabs + projct*(apec*tbabs) +  gauss} (1.49 keV)\textit{ + gauss} (1.75 keV) and \textit{apec+(apec+apec+pegpwrlw)*tbabs + projct*(apec*tbabs)} for \textit{XMM-Newton} and \textit{Chandra} data respectively. The XRB model normalisations and cluster abundance (Z) within each region were fixed to values obtained from the projected analysis. The regions used for spectral extraction and the temperature (kT) and electron density (n$_{e}$) values are listed. The errors are quoted at 90 per cent confidence level.}
  \label{tab:radial_prof_projected}
  \begin{tabular}{ccccc}
    \hline
\hline
X-ray group &Region&Data & kT & n$_{e}$\\
&&&(keV)&($10^{-3}$cm$^{-3}$)\\
\\
\hline
\hline
A2151C(B)&0--40 arcsec&\textit{XMM-Newton}&$1.64_{-0.06}^{+0.08}$&$8.51_{-0.25}^{+0.24}$\\
& &\textit{Chandra}&$1.91\pm{0.07}$&$10.64_{-0.16}^{+0.15}$\\
\\
&40--80 arcsec&\textit{XMM-Newton}&$2.19_{-0.19}^{+0.24}$&$4.23_{-0.12}^{+0.11}$\\
& &\textit{Chandra}&$2.38_{-0.15}^{+0.18}$&$4.55\pm{0.09}$\\
\\
&80--120 arcsec&\textit{XMM-Newton}&$2.01_{-0.18}^{+0.25}$&$2.78_{-0.10}^{+0.08}$\\
& &\textit{Chandra}&$2.42_{-0.21}^{+0.25}$&$3.06_{-0.04}^{+0.08}$\\
\\
&120--160 arcsec&\textit{XMM-Newton}&$2.82_{-0.54}^{+0.60}$&$2.18\pm{0.08}$\\
& &\textit{Chandra}&$2.50_{-0.30}^{+0.41}$&$2.02\pm{0.08}$\\
\\
&160--210 arcsec&\textit{XMM-Newton}&$2.37_{-0.45}^{+0.65}$&$1.58_{-0.07}^{+0.08}$\\
& &\textit{Chandra}&$1.93_{-0.09}^{+0.10}$&$2.24\pm{0.03}$\\
\\
&210--300 arcsec (280\textdegree{} wide sector)&\textit{XMM-Newton}&$1.85_{-0.16}^{+0.18}$&$1.42_{-0.12}^{+0.03}$\\
& &\textit{Chandra}&-&-\\
\hline
A2151C(F)&0--50 arcsec&\textit{XMM-Newton}&$1.07_{-0.05}^{+0.06}$&$3.87\pm{0.15}$\\
\\
&50--100 arcsec&&$1.04_{-0.08}^{+0.09}$&$1.29_{-0.11}^{+0.10}$\\
\\
&100--150 arcsec&&$2.45_{-0.87}^{+1.70}$&$1.07_{-0.11}^{+0.10}$\\
\\
&150--210 arcsec (230\textdegree{} wide sector)&&$1.09_{-0.08}^{+0.11}$&$1.02\pm{0.08}$\\
\\
&210--280 arcsec (230\textdegree{} wide sector)&&$1.22_{-0.10}^{+0.15}$&$0.72_{-0.07}^{+0.06}$\\
\\
&280--380 arcsec (230\textdegree{} wide sector)&&$2.21_{-1.01}^{+3.16}$&$0.46_{-0.08}^{+0.09}$\\
\\
&380--480 arcsec (230\textdegree{} wide sector)&&$1.08_{-0.08}^{+0.10}$&$0.89\pm{0.03}$\\
\hline\hline
  \end{tabular}
  \label{tab:tab4}
  \end{center}
 \end{table*}

\begin{figure}
\centering
\includegraphics[width=0.48\textwidth,height=6.1cm]{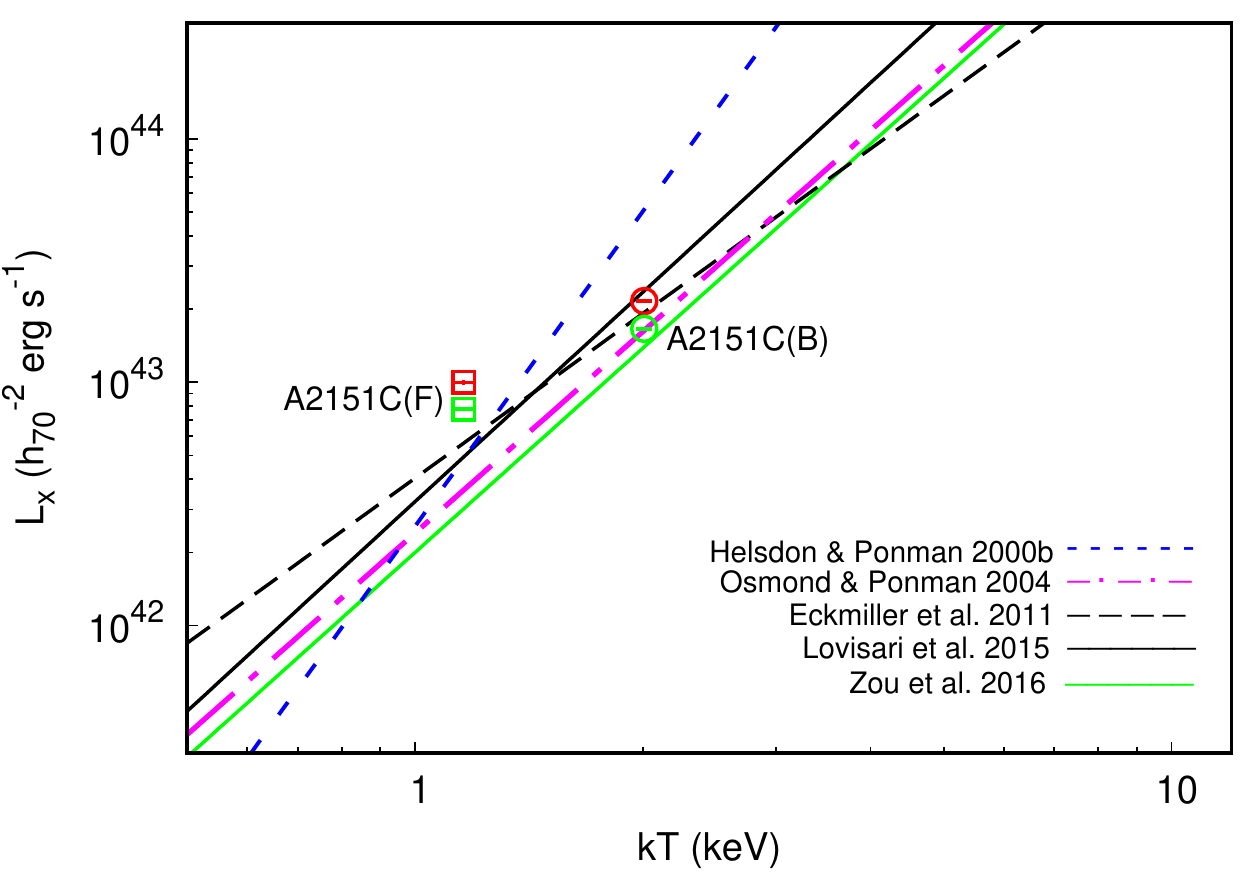}
\caption{\textit{L$_{X}$-kT} relations of galaxy groups obtained by different authors. The \textit{open circle} and \textit{open square} symbols mark the positions of A2151C(B) and A2151C(F) respectively, with the X-ray luminosity estimated in two energy bands $-$ $0.1-2.4$ keV (red symbols; for comparison with all studies except \citet{zou2016}) and $0.5-2.0$ keV (green symbols; for comparison with \citet{zou2016}). The errors in $L_{X}$ and $kT$ are of the size of the symbols or smaller.}
\label{fig:fig5}
\end{figure}

\subsection{X-ray Luminosity}
X-ray luminosities ($L_X$) of A2151C(B) and A2151C(F) were estimated by convolving the cluster \textit{apec} model component used in the spectral analysis in \S3.2 and \S3.3.1 with the XSPEC model \textit{clumin} for both \textit{XMM-Newton} and \textit{Chandra} data, after freezing the cluster \textit{apec} normalisation. The global spectral analysis of the two subclumps made use of regions which were restricted to sectors in the outer parts for reasons stated in \S3.2. Luminosities of the excluded regions were estimated from the spectral analysis of the outer sectors assuming that each smaller region of equal area within an annulus contributes equally to the total annular luminosity. These values were then added to the luminosity values obtained from the global spectral analysis. As an example, we describe the method used to estimate the total X-ray luminosity of A2151C(B). The X-ray luminosity of the 280\textdegree{} wide sector with inner and outer radii of 210 and 300 arcsec respectively (outermost region of A2151C(B) in \autoref{fig:fig2}\textcolor{blue}{a}; $L_{X\_280}$) was estimated from the spectral analysis of this region. The luminosity of the missing 80\textdegree{} wide sector ($L_{X\_80}$), with inner and outer radii of 210 and 300 arcsec respectively (excluded from the global spectral analysis) was then estimated as follows: \\\\$L_{X\_80}=\frac{80}{280} \times L_{X\_280}$

This estimate was added to the value obtained from the global spectral analysis ($L_{X\_global}$) in order to obtain the total X-ray luminosity ($L_{X\_total}$) of A2151C(B):\\\\$L_{X\_total} = L_{X\_global} +  L_{X\_80}$

The total luminosity of A2151C(F) was estimated in the same manner. The total X-ray luminosities of the two subclumps are listed in \autoref{tab:tab2}.  Errors on the total luminosity estimates were obtained using the standard error propagation technique. A2151C(B) is $\sim$3 times more luminous than A2151C(F).

We also obtained  X-ray luminosities of the two subclumps in the energy bands 0.1$-$2.4 keV and 0.5$-$2.0 keV using \textit{XMM-Newton} data to see where they are located with respect to the \textit{L$_X$-kT} relation obtained for groups of galaxies by several authors using the above energy bands. These relations and the associated references are given in \autoref{tab:tab6}. In order to obtain the temperature and luminosity of galaxy groups used in deriving the \textit{L$_X$-kT} relations, all authors except \citet{zou2016} performed a global spectral analysis within the entire fov or used a circular region of radius \textit{R$_{spec}$}, which is the extent up to which X-ray emission from the group was detected. \citet{zou2016} estimated the ICM properties within the $r_{500}$ aperture.  

\autoref{fig:fig5} shows these relations along with the positions of A2151C(B) and A2151C(F) (their $L_X$ values scaled by $h_{70}^{-2}$ for consistency). A2151C(B) follows the group \textit{L$_X$-kT} relations very closely, thus, suggesting that its properties are very similar to those of galaxy groups. A2151C(F) lies close to the \textit{L$_X$-kT} relations but with a slightly higher X-ray luminosity. We note that the global spectral analysis of A2151C(F) was performed within a region of radius 480 arcsec since significant emission was detected up to this radius (\S3.2). The subclump is however fairly regular only up to $\sim$300 arcsec. Including regions beyond this radius in the global spectral analysis, which may not necessarily be part of the subclump and might correspond to infalling material, may have led to a slight underestimation of temperature and overestimation of luminosity.
\begin{table*}
 \begin{center}
  \captionsetup{justification=centering}
  \caption{Mass of the intragroup gas obtained by fitting single-$\beta$ models to the projected gas density profiles of A2151C(B) and A2151C(F). The errors are quoted at 90 per cent confidence level.}
  \label{tab:global}
  \begin{tabular}{cccccccccc}
    \hline
\hline
Subclump & Data & $\beta$ & $r_{c}$  & $\rho_{0}$ &$(\chi_{\nu}^2)_{\text{min}}$ & $r$ & $M_{gas}(r)$ & $r_{500}$ &$M_{gas\_500}$   \\
&&&(kpc)&$(10^{4} M_{\odot} \text{ kpc}^{-3})$&&(kpc)&$(10^{12} M_{\odot})$ & (kpc) & $(10^{12} M_{\odot})$ \\
\hline
A2151C(B)&\textit{XMM-Newton}&$0.38\pm{0.03}$&$19.35_{-3.91}^{+4.04}$&$23.40_{-2.42}^{+3.23}$&1.46&304&$1.89_{-0.84}^{+1.31}$ &$830.95$ & $12.44^{+9.66}_{-6.11}$ \\
&\textit{Chandra}&$0.38\pm{0.03}$&$16.63_{-3.85}^{+3.78}$&$28.07_{-3.44}^{+5.57}$&1.01&304&$1.92_{-0.93}^{+1.60}$&$830.95$ &$12.60^{+8.71}_{-6.70}$ \\
\\
\hline
A2151C(F)&\textit{XMM-Newton}&$0.28\pm{0.02}$&$10.43_{-1.49}^{+1.59}$&$13.20\pm{0.44}$&1.93&326.8&$1.48_{-0.44}^{+0.56}$&$629.30$&$6.11^{+2.66}_{-1.97}$ \\
&&&&&&304&$1.27_{-0.37}^{+0.47}$&\\
\hline\hline
  \end{tabular}
  \label{tab:tab5}
  \end{center}
 \end{table*}

\subsection{Gas Mass Estimates}\label{sec3.5}
We estimated the gas mass of A2151C(B) and A2151C(F) by using the gas density profiles obtained in \S3.3.1. The projected gas density profiles of the two subclumps were fitted using a single-$\beta$ model given by:
\begin{equation} n_{e}(r)=n_{e}(0)\left(1+\frac{r^2}{r_c{^2}}\right)^\frac{-3\beta}{2}\label{eq2}\end{equation}
where $n_{e}(0)$ and $r_{c}$ are the central gas density and projected core radius respectively, and $\beta$ is the ratio of the specific energy in galaxies to the specific energy in hot gas.
We also tried fitting a double-$\beta$ model to the gas density profiles of A2151C(B) and A2151C(F) but this did not give significant results due to very few data points available.

The gas mass $M_{gas}(R)$ within a certain radius R can be obtained by using the following:
\begin{equation} M_{gas}(R)=4\pi \rho_{0}\int\displaylimits_0^R\left(1+\frac{r^2}{r_c{^2}}\right)^\frac{-3\beta}{2} r^{2} dr\label{eq3}\end{equation}
where $\rho_{0}=\mu n_{e}(0) m_{p}$, $m_{p}=1.67\times10^{-27} \text{ kg}$ is the mass of proton and $\mu=0.613$ is the mean molecular weight for a fully ionized gas (derived for an ICM abundance of 0.43 Z$_{\odot}$ using the abundance table of \citet{wilms2001}; \S3.3.1). We obtained the gas mass of A2151C(B) and A2151C(F) out to radii 400 arcsec ($\sim$304 kpc) and 430 arcsec ($\sim$326.8 kpc) respectively from the corresponding X-ray peaks. The results of the single-$\beta$ model fitting to the projected gas density profiles along with the gas mass estimates are listed in \autoref{tab:tab5}.

We believe that the gas mass value for A2151C(F) may have been overestimated. Fig. \ref{fig:fig1}\textcolor{blue}{(a)} shows that the X-ray emission of A2151C(F) is fairly regular only up to $\sim$300 arcsec. Beyond 300 arcsec radius, several other smaller infalling groups may be merging with the main A2151C(F) group (see Fig. \ref{fig:fig1}\textcolor{blue}{(b)} and \S4). The actual extent of A2151C(F) may therefore be $\sim$300 arcsec which is smaller than the 400 or 430 arcsec radii within which the gas mass values are reported in \autoref{tab:tab5}. We also note that spherical symmetry of A2151C(F) has been assumed while calculating the gas mass which may not be a reasonable assumption for radii beyond 300 arcsec. Moreover, the $\beta$ value obtained for A2151C(F) which has been used in the gas mass estimation is unusually low (discussed further in this subsection). For comparison, we also calculated the gas mass within 300 arcsec ($\sim$228 kpc) radius of each of the two groups which resulted in a value of $1.10^{+0.71}_{-0.47}\times10^{12} M_{\odot}$ for A2151C(B) and $6.79^{+2.40}_{-1.90}\times10^{11} M_{\odot}$ for A2151C(F). These values indicate that the gas mass within 300 arcsec of A2151C(B) is greater than that of A2151C(F), although the errors are large.
We also obtained the gas mass of each subcluster within a projected radius $r_{500}$, the radius from the adopted cluster centre at which the total mass density is 500 times the critical density of the Universe. $r_{500}$ was estimated using the following relation derived from simulations by \citet{evrard1996} and adopted by \citet{osmond2004} : \begin{equation} r_{500}=\frac{124}{H_0}\sqrt{\frac{kT}{10 \text{ keV}}} \text{ Mpc},\end{equation} where $kT$ is the gas temperature in keV obtained from the global spectral analysis of the cluster within a radius up to which the X-ray emission is detected and $H_{0}$ is the Hubble constant in km s$^{-1}$ Mpc$^{-1}$. The resulting $r_{500}$ value of each subcluster along with the gas mass estimates within this radius ($M_{gas\_500}$) are provided in \autoref{tab:tab5}. We note that for obtaining the $r_{500}$ value of A2151C(B), an average of the temperature values (2.04 keV) obtained from the \textit{XMM-Newton} and \textit{Chandra} global spectral analysis was used.

The $\beta$ value obtained for A2151C(B) is similar to that found in groups of galaxies ($\lesssim 0.5$ \citet{helsdon2000,mulchaey2000}) but that obtained for A2151C(F) is unusually low. Both $\beta$ and core radius values of A2151C(B) are comparable to those obtained by \citet{huangsarazin96}. We also compared the $\beta$ value resulting from fitting the gas density profile ($\beta_{fit}$) to that obtained from \begin{equation} \beta=\frac{\mu m_{p} \sigma^{2}}{kT}\label{eq4}\end{equation} where $\sigma$ is the measured optical velocity dispersion of the subclump and $kT$ is the spectrally determined temperature of the intragroup gas. In case of A2151C(B), equation \eqref{eq4} resulted in a $\beta$ value $\sim$0.6 using $\sigma=441$ km s$^{-1}$ from \citet{agulli2017}. This value is quite close to that obtained from fitting the single-$\beta$ model to the gas density profile. \citet{huangsarazin96} arrived at $\beta$ $\sim$1.87 on application of eq. \eqref{eq4} and $\beta_{fit}$ $\sim$0.45 resulting from fitting the $\beta$ model to the surface brightness profile of A2151C(B). They interpreted this large discrepancy between the two $\beta$ values as a consequence of the dynamically unrelaxed status of the system. This discrepancy seems to be resolved here. In case of A2151C(F), however, eq. \eqref{eq4} gave $\beta$ $\sim$3 with $\sigma=711$ km s$^{-1}$ \citep{agulli2017} which is not at all in agreement with the $\beta_{fit}$ value obtained from the gas density profile fitting ($\sim$0.28). This disagreement may be due to two reasons -- either the value of $\sigma$ for A2151C(F) has been overestimated or A2151C(F) is a dynamically unrelaxed subclump which is itself composed of several small groups of galaxies as can be seen in Fig. \ref{fig:fig1}\textcolor{blue}{(b)}.

We did not use the results of the deprojection analysis to estimate the gas mass due to the large error bars obtained in the best-fitting parameter values.

 \begin{table*}
 \begin{center}
  \captionsetup{justification=centering}
  \caption{Best-fitting \textit{L$_X$-kT} relations obtained by different authors for groups of galaxies}
  \begin{tabular}{cccc}
    \hline
    \hline
Author&Relation\textsuperscript{a}&X-ray Data&Group Sample Size\\\hline
\citet{helsdon2000b} (fig.1, table 2)&log $(L_X/h_{50}^{-2} \text{erg s}^{-1}) = 4.30$ log $(kT/1\text{ keV})$ + $42.70$&\textit{ROSAT}&42\\
\citet{osmond2004} (fig. 12, table 6)&log $(L_X/h_{70}^{-2} \text{erg s}^{-1}) = 2.75$ log $(kT/1\text{ keV})$ + $42.38$&\textit{ROSAT}&35 (G-sample)\textsuperscript{ b}\\
\citet{eckmiller2011}\textsuperscript{ c} (fig.3, table 4)&log $(L_X/0.5 \times 10^{44} h_{70}^{-2} \text{erg s}^{-1}) = 2.25$ log $(kT/3\text{ keV}) -0.02$&\textit{Chandra}&26\\
\citet{lovisari2015} (fig. 2, tables 3 and 4)&log $(L_X/10^{43} h_{70}^{-2} \text{erg s}^{-1}) = 2.86$ log $(kT/2\text{ keV}) + 0.37$\textsuperscript{ d}&\textit{XMM-Newton}&20\\
\citet{zou2016} (fig. 3, sec. 4.3)&log $(L_X/10^{43} h_{70}^{-2} \text{erg s}^{-1}) = 2.79$ log $(kT/2\text{ keV}) + 0.14$\textsuperscript{d}&\textit{Chandra}&23\\
\hline\hline \end{tabular}
  \label{tab:tab6}
  \end{center}
\small\textsuperscript{a}The X-ray luminosity values used are estimated in the energy range $0.1-2.4$ keV except in \citet{zou2016} who used the soft X-ray band ($0.5-2.0$ keV) to estimate the luminosity values. 
\\\small\textsuperscript{b} Systems that have group-scale emission.
\\\small\textsuperscript{c}The X-ray luminosity values in the energy range $0.1-2.4$ keV used by the authors were all taken from the input catalogs and were determined homogeneously from \textit{ROSAT} observations. 
\\\small\textsuperscript{d}This relation by the authors is corrected for the selection bias effects. 
 \end{table*}
 
\subsection{Total Mass Estimates}\label{sec3.6}
The total gravitational mass of a cluster/group within radius $r$, $M_{grav}(<r)$, under the assumption of hydrostatic equilibrium can be calculated by using the following:
 \begin{equation}
M_{grav}(<r)=\frac{-rkT(r)}{G\mu m_{p}}\Bigg[\frac{dlnT(r)}{dlnr}+\frac{dlnn_{e}(r)}{dlnr}\Bigg]\label{eq5}     
 \end{equation}
where $G$ is the gravitational constant and the other symbols have the same meaning as defined previously. Using equation \eqref{eq5} we estimated the total gravitational mass of A2151C(B) within a radius of 304 kpc ($\sim$400 arcsec) from the X-ray peak to be $1.60\pm{0.40}\times10^{13}$ M$_{\odot}$. The spectral analysis with \textit{XMM-Newton} data allowed determination of gas temperature and density up to a radius of 300 arcsec (228 kpc). In order to obtain the total gravitational mass within the radius up to which X-ray emission is seen in A2151C(B) ($\sim$400 arcsec; 304 kpc; Fig. \ref{fig:fig1}), we assumed the gas temperature and density at this radius to be equal to the best-fitting values obtained for the outermost annular sector (210$-$300 arcsec; Fig. \ref{fig:fig2}\textcolor{blue}{(a)}; Table \ref{tab:tab3}) in the spectral analysis. The mass estimate based on the \textit{XMM-Newton} data shows that the gas mass is $\sim$12 per cent of the total mass. Calculations by \citet{huangsarazin96} show a similar value for the gas mass fraction at 400 arcsec ($\sim$15 per cent) in A2151C(B).
  
The gas temperature and density in the outermost annular sector (210$-$300 arcsec) could not be determined with \textit{Chandra} data due to poor statistics (\S3.3.1). Hence, \textit{Chandra} data could only give a total mass estimate for a smaller radius. The total mass of A2151C(F) within a radius of 326.8 kpc ($\sim$430 arcsec) was calculated to be $1.44\pm{0.62}\times 10^{13}$ M$_{\odot}$ using \textit{XMM-Newton} data. We also estimated the total gravitational mass of the two subclumps within $r_{500}$ ($M_{grav\_500}$) by extrapolating the total mass values obtained at smaller radii. The $M_{grav\_500}$ values are $9.08\pm{5.24}\times 10^{13}$ M$_{\odot}$ and $3.01\pm{2.11}\times 10^{13}$ M$_{\odot}$  for A2151C(B) and A2151C(F) respectively. We note that the results of the projected spectral analysis were used in the above calculations. The error bars on the mass values correspond to a 90 per cent confidence interval.

\subsection{Cooling time}\label{sec3.7}
\begin{figure*}
\begin{multicols}{2}
\subcaptionbox{}{\includegraphics[width=0.92\linewidth,height=5.9cm]{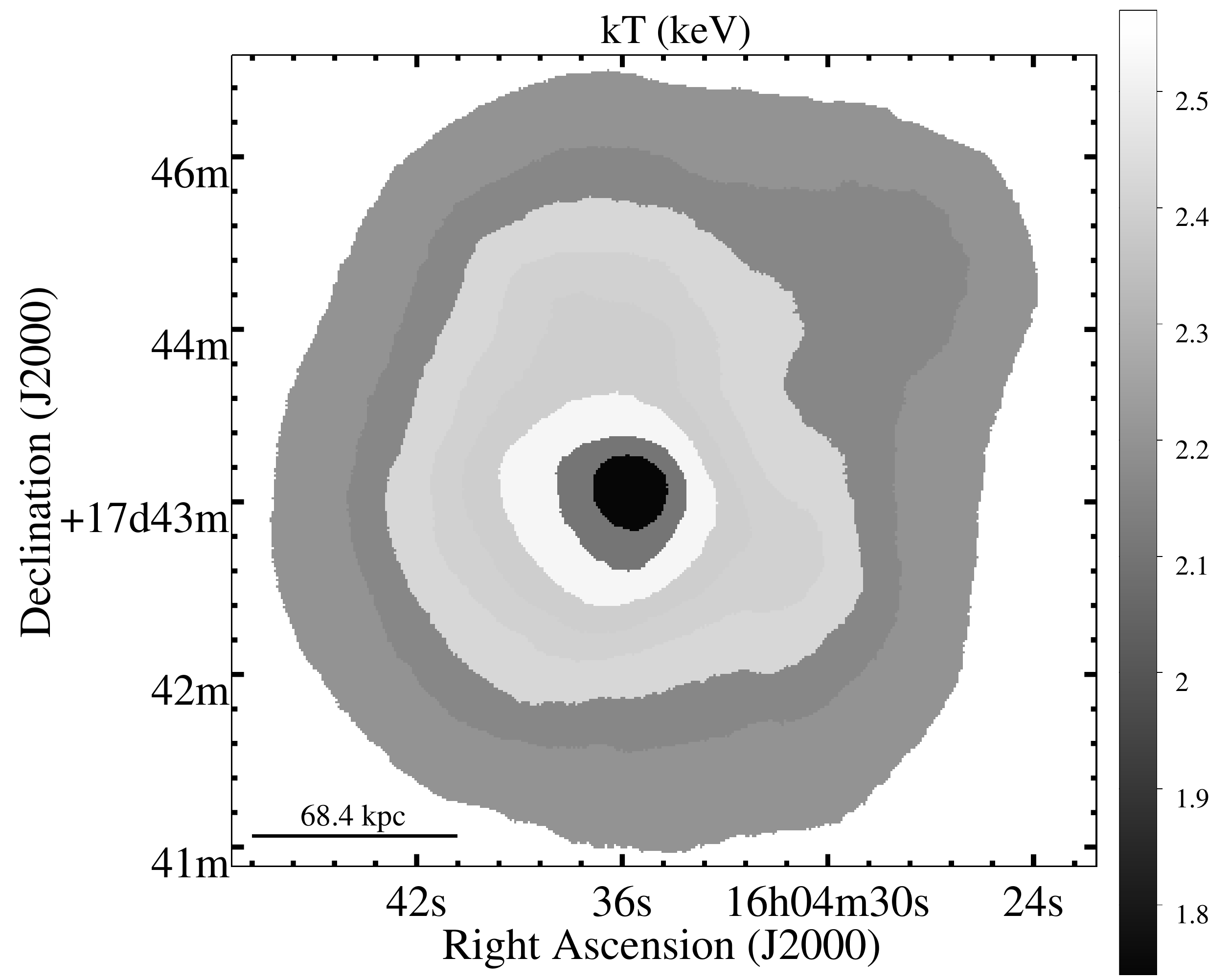}}\par
\subcaptionbox{}{\includegraphics[width=0.92\linewidth,height=5.9cm]{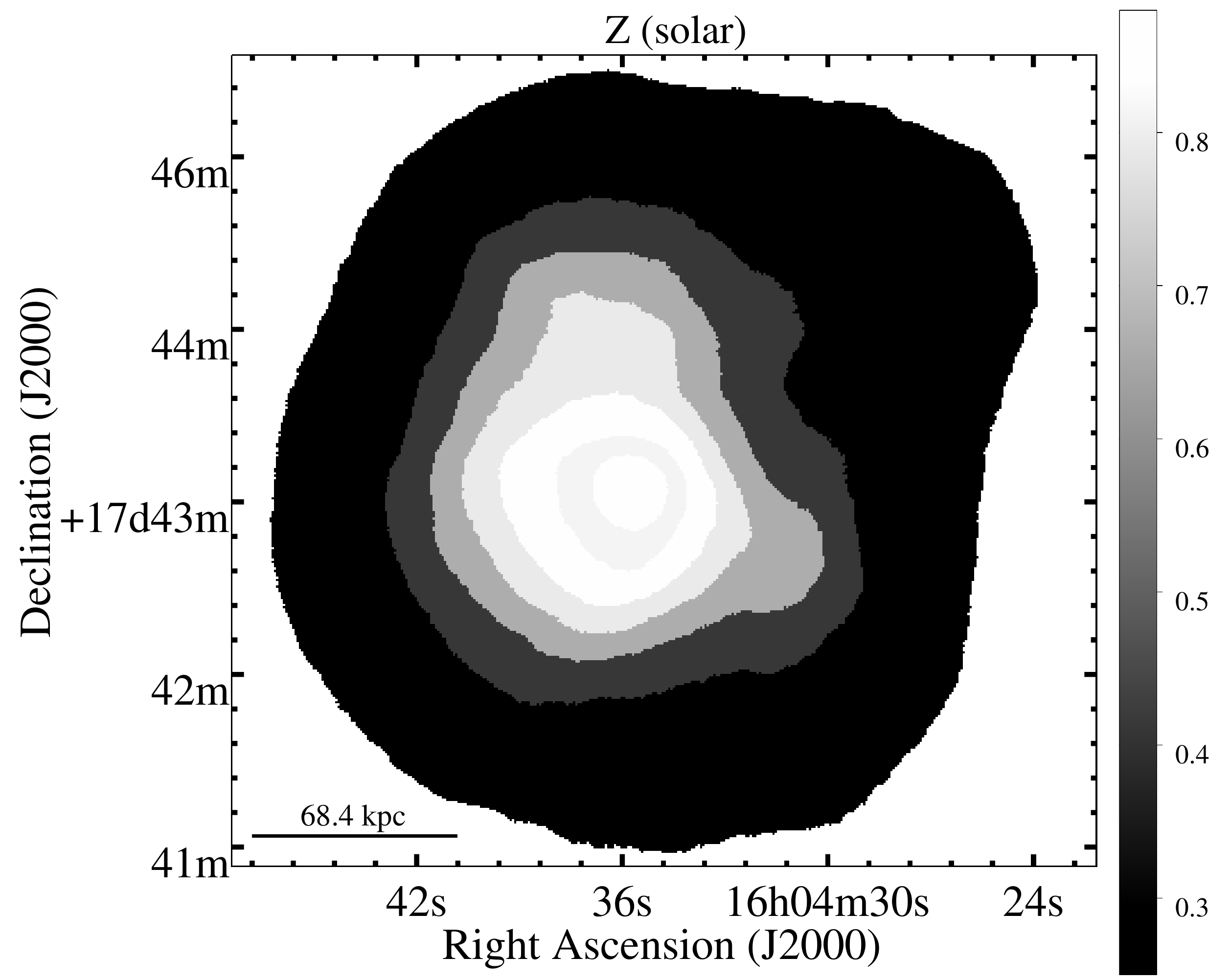}}\par
\end{multicols}
\begin{multicols}{2}
\subcaptionbox{}{\includegraphics[width=0.92\linewidth,height=5.9cm]{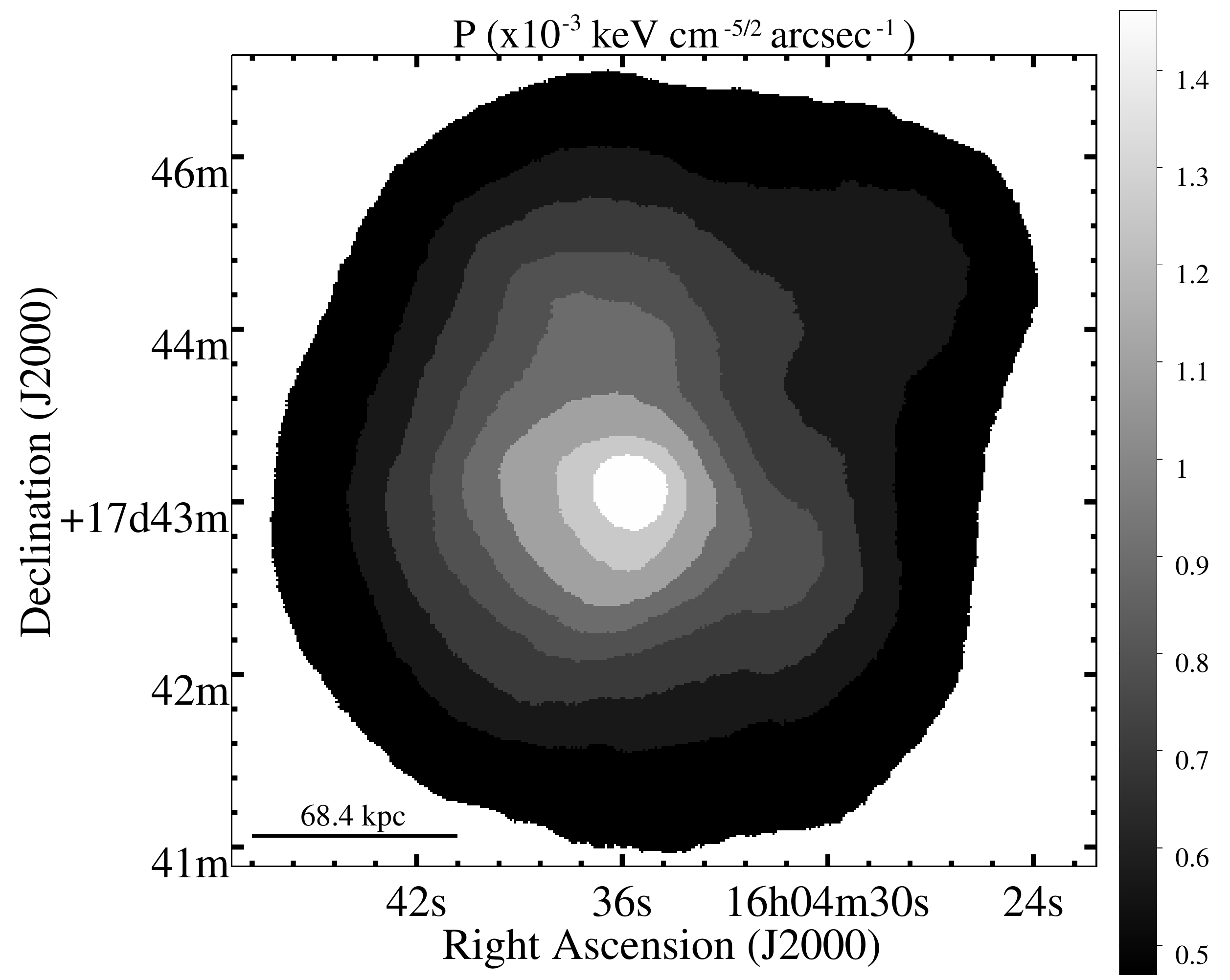}}\par
\subcaptionbox{}{\includegraphics[width=0.92\linewidth,height=5.9cm]{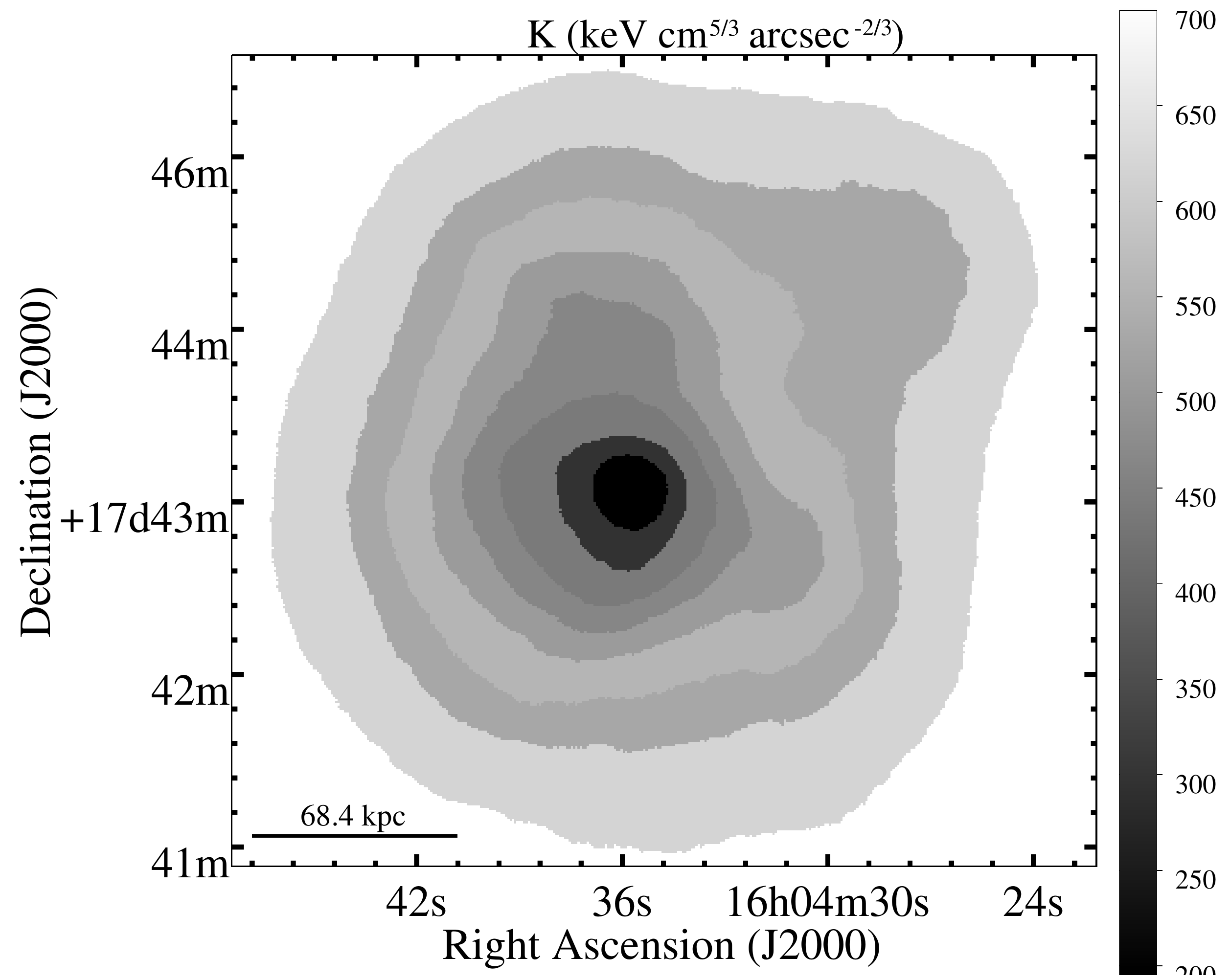}}\par
\end{multicols}
\caption{2D maps of A2151C(B) obtained as a result of the application of \textit{contbin} algorithm to \textit{Chandra} data. Panel (a)--(b): Projected temperature (kT) and abundance (Z) obtained directly from spectral fitting; (c)--(d): Projected pressure (P) and entropy (K).}
\label{fig:fig6}
\end{figure*}

The thermodynamic maps of A2151C(B) obtained from the spectral analysis of Chandra data (\S3.8.1), show a low temperature core of radius $\sim$15 arcsec (\autoref{fig:fig6}). We performed a deprojected spectral analysis using Chandra spectra extracted from six annular regions with outer radii 15, 50, 80, 120, 160, and 210 arcsec. The resulting deprojected temperature and electron number density values within the central 15 arcsec ($\sim$11 kpc) circular region of A2151C(B) are $1.55\pm{0.07}$ keV and $2.54^{+0.06}_{-0.07} \times 10^{-2} \text{cm}^{-3}$ respectively. Using these values, we calculated the cooling time of A2151C(B) in the central core using the following relation from \citet{panagoulia2014} :
\begin{equation}
\begin{split}
t_{cool}=0.8\times10^{10} \text{ yr}\left(\frac{n_{e}}{10^{-3} \text{cm}^{-3}}\right)^{-1} \left(\frac{\text{T}}{10^{7}\text{ K}}\right)^{1.6}\\
\text{ for T}< 3 \times 10^{7} \text{ K}
\end{split}
\label{eq6}
\end{equation}
This resulted in a cooling time of $8.05_{-0.76}^{+0.84} \times 10^{8} \text{ yr}$.
 The estimated cooling time is much smaller than the Hubble time ($\sim$14.5 $\times 10^{9} \text{ yr}$ in our cosmology), thus confirming the presence of a cool core in A2151C(B).
 
\subsection{Maps of the ICM thermodynamical quantities}
\subsubsection{Chandra data: Contour binning maps}It can be seen from Fig. \ref{fig:fig1} that X-ray emission in A2151C(B) and A2151C(F) is not precisely azimuthally symmetric. Azimuthally averaged spectral analysis, although insightful, may therefore not be the best way to study the variation of parameters of the intracluster medium (e.g. temperature and abundance). Moreover, these parameters usually change in the direction of surface brightness changes \citep{sanders2006}.  We implemented the `contour binning' algorithm of \citet{sanders2006} which uses the X-ray surface brightness distribution to define spatial bins which cover regions of similar brightness and then performed spectral fitting in each of these bins. Since the technique is designed to work with \textit{Chandra} data, we could obtain the thermodynamic (TD) maps using this algorithm for A2151C(B) only (A2151C(F) is only partially covered by \textit{Chandra} observations). The point-source-subtracted, hole-filled image of the observation with the highest exposure time (ObsID 20087) was used to generate spatial bins. It was accumulatively smoothed with a signal-to-noise ratio (SNR) threshold of 25 and then binned with a minimum SNR of 38. An exposure map and a scaled particle-background image were also supplied to the task \textit{contbin}.   Eight spatial bins were obtained. Spectra and responses were extracted in each of these bins for the three \textit{Chandra} observations and fitted simultaneously using the model \textit{constant*(apec+(apec+apec+pegpwrlw)*tbabs + apec*tbabs)}. Temperature and abundance maps were generated using the best-fitting values obtained as a result of spectral fitting.

The maps of these quantities are shown in \autoref{fig:fig6}.  A drop in the temperature towards the centre of A2151C(B) seen in Fig. \ref{fig:fig6}\textcolor{blue}{(a)} is consistent with the presence of a cool core which was also observed in the temperature profile in Fig. \ref{fig:fig3}\textcolor{blue}{(a)}.   As we go outwards from this central region, the temperature increases and then again begins to drop in the outskirts of the subclump.   The typical error (at 90 per cent confidence level) in the temperature value is $\sim$0.1-0.2 keV.    Fig. \ref{fig:fig6}\textcolor{blue}{(b)} indicates that the central regions of the subclump have a higher metallicity as compared to the metallicity in the outer regions. The abundance values, however, are not very well constrained.   Typical errors (at 90 per cent confidence level) are $\sim$0.1-0.3 Z$_\odot$. In order to obtain the projected pressure (P) and entropy (K) maps, we made use of the cluster \textit{apec} normalisation $\mathcal{N}$ (\autoref{eq:eq1} in \S3.3.1) and the projected emission measure \textit{EM} defined as follows.
\begin{equation}\textit{EM}=\mathcal{N}/\textit{A}\label{eq7} \end{equation} where \textit{A} is the area of the spatial bin.\\The projected emission measure \textit{EM} is thus, proportional to the square of the electron density integrated along the LOS. Using \eqref{eq7}, the projected pressure can be computed as
\begin{equation}P=kT (EM)^{1/2} \hspace{0.6cm} (keV cm^{-5/2} arcsec^{-1})\label{eq8} \end{equation} and projected entropy as
\begin{equation} K=kT (EM)^{-1/3} \hspace{0.4cm} (keV cm^{5/3} arcsec^{-2/3})\label{eq9} \end{equation}
A similar method has been adopted in \citet{botteon2018}.

The 2D maps of pressure and entropy are displayed in \autoref{fig:fig6}\textcolor{blue}{(c)} and \ref{fig:fig6}\textcolor{blue}{(d)} respectively.

 \begin{figure*}
\begin{multicols}{2}
\subcaptionbox{}{\includegraphics[width=\linewidth,height=5.2cm]{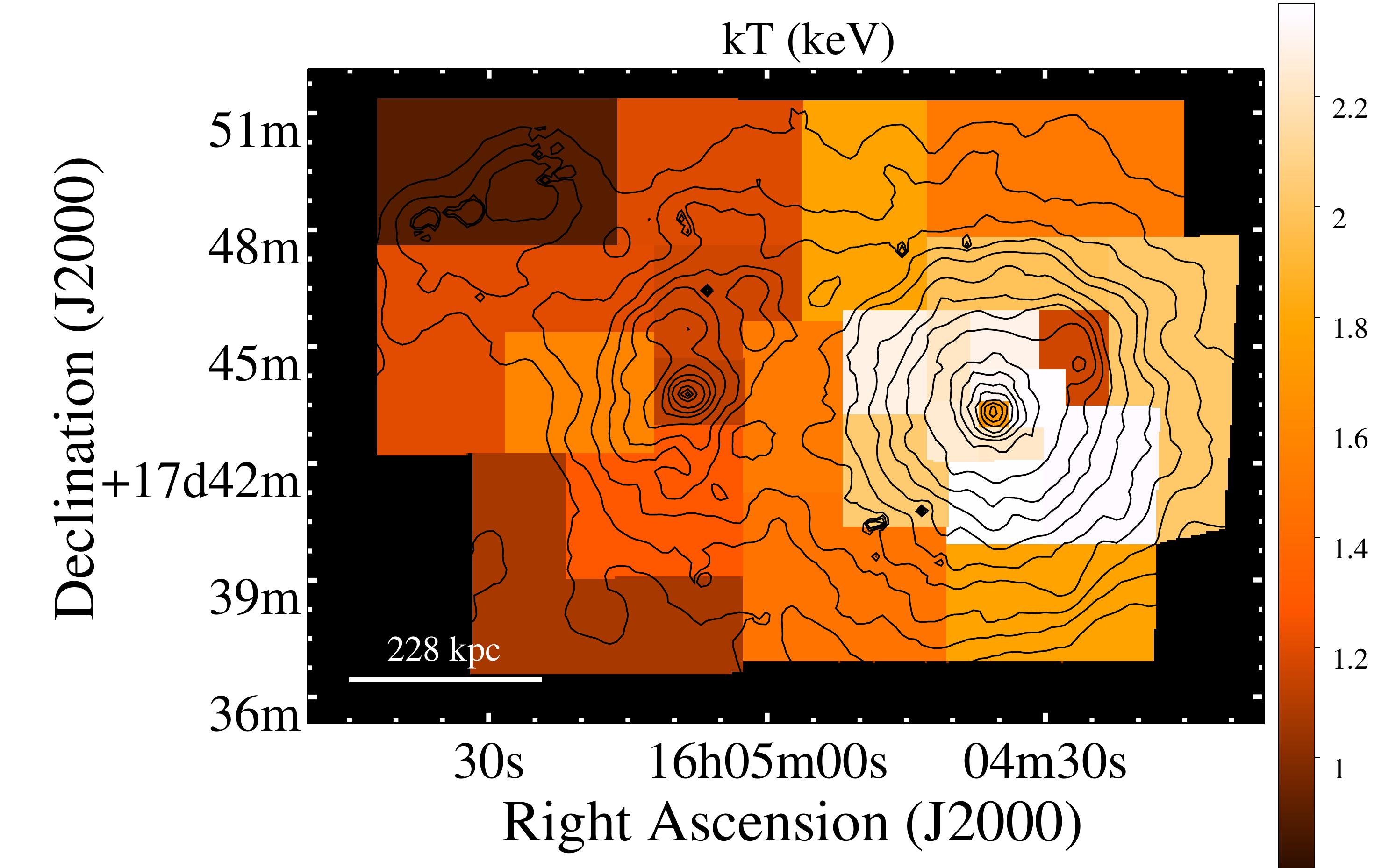}}\par
\subcaptionbox{}{\includegraphics[width=\linewidth,height=5.2cm]{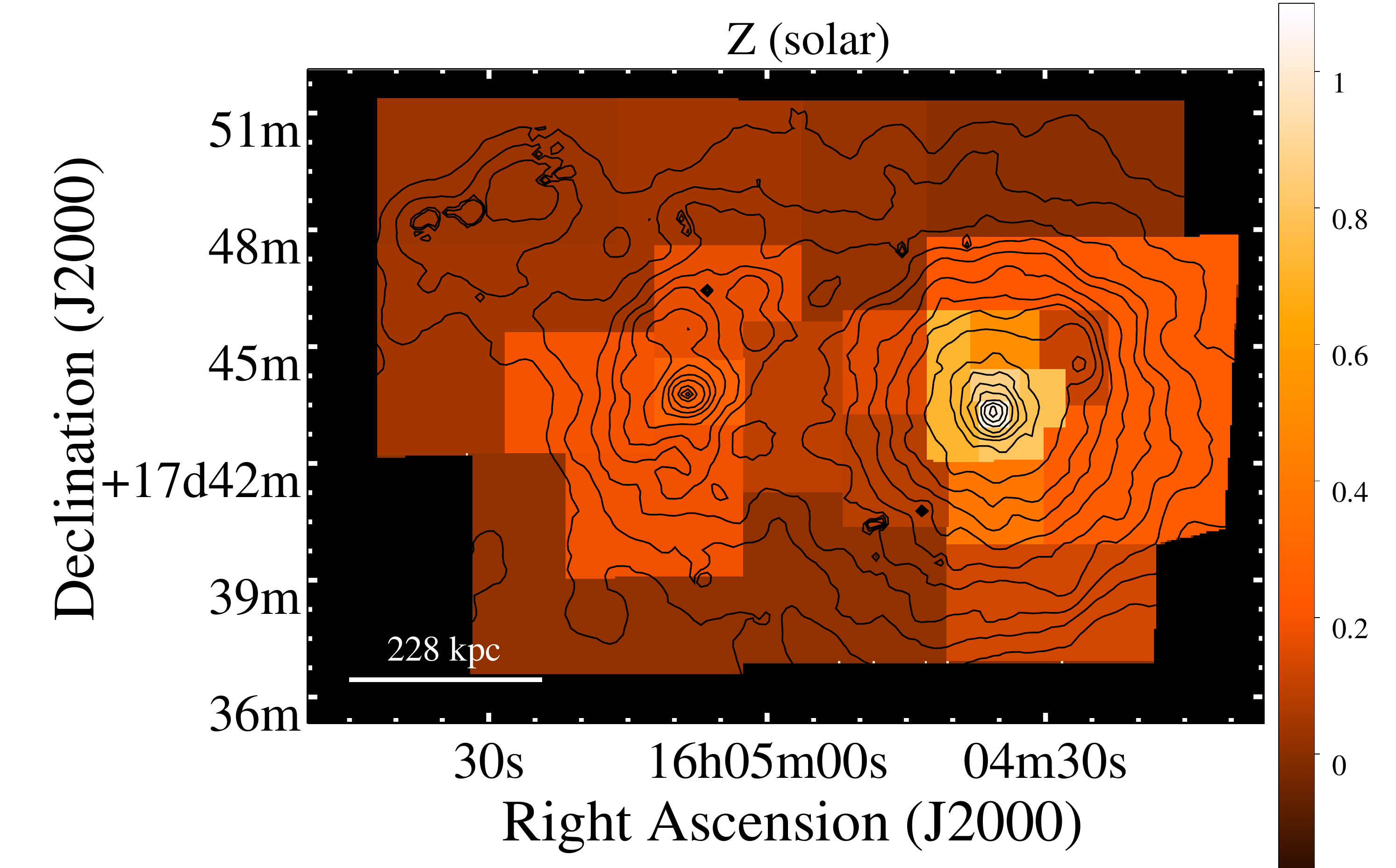}}\par
\end{multicols}
\begin{multicols}{2}
\subcaptionbox{}{\includegraphics[width=\linewidth,height=5.2cm]{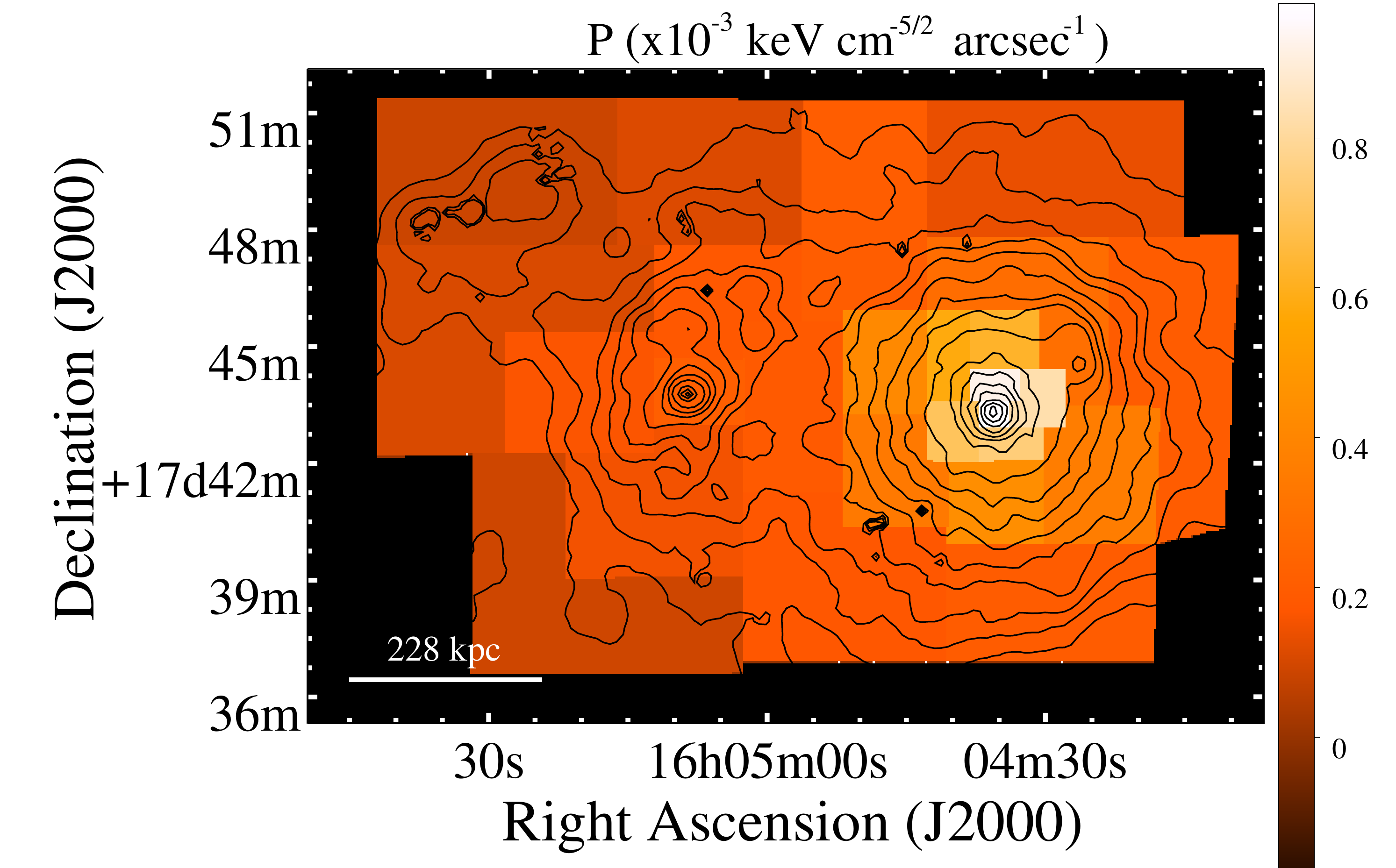}}\par
\subcaptionbox{}{\includegraphics[width=\linewidth,height=5.2cm]{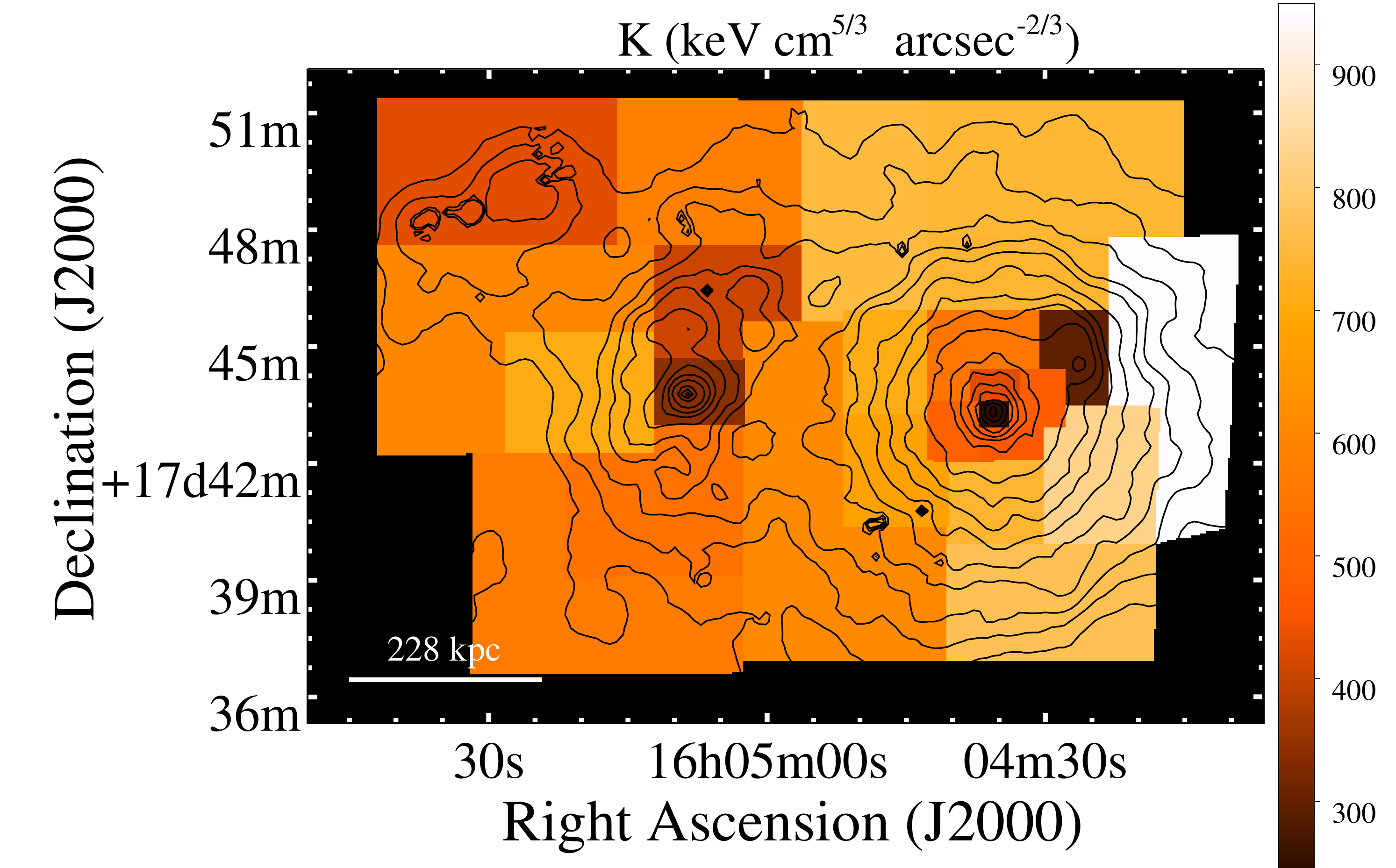}}\par
\end{multicols}
\caption{Panel (a)--(d): Projected maps of temperature (kT), abundance (Z), pressure (P) and entropy (K) respectively obtained from the analysis of spectra extracted in 27 box/polygon shaped regions using \textit{XMM-Newton} data.}
\label{fig:fig7}
\end{figure*}

\subsubsection{XMM-Newton data}
To study the 2D spatial distribution of the thermodynamic properties and metallicity of the ICM using \textit{XMM-Newton} data, we extracted spectra from 27 box/polygon shaped regions in each of the three detectors. The regions were chosen to have a minimum of 600 counts in each detector. The extracted PN, MOS1 and MOS2 spectra were then suitably grouped and fitted simultaneously with \textit{constant*(apec+(apec+apec+pegpwrlw)*tbabs + apec*tbabs) + gauss} (1.49 keV) \textit{+ gauss }(1.75 keV) model in the energy range 0.3$-$7.0 keV. 2 out of 27 regions were not covered by the PN detector or almost completely affected by its chip gaps. These were analysed using MOS1 and MOS2 data only. One region which overlapped with the pair of galaxies NGC 6040 and PGC 56942 (the rightmost red box in \autoref{fig:fig1}\textcolor{blue}{(b)}), was fitted with the model \textit{constant*(apec+(apec+apec+pegpwrlw)*tbabs + (apec+powerlaw)*tbabs) + gauss} (1.49 keV) \textit{+ gauss} (1.75 keV). The \textit{powerlaw} component was added here to account for the AGN emission from NGC 6040 \citep{best2012} which was not subtracted out in the point source detection and removal procedure carried out on the \textit{XMM-Newton} data. Temperature and abundance values for each region were obtained directly as a result of the fit. We used equations \eqref{eq7} - \eqref{eq9} to derive the projected pressure and entropy in each region.

The projected maps were constructed using the FTOOL \textit{ftimgcalc}. These are displayed in \autoref{fig:fig7}. Fig. \ref{fig:fig7}\textcolor{blue}{(a)} shows the 2D temperature map of A2151C(B) and A2151C(F). The temperature distribution does not appear to be azimuthally symmetric. A  temperature of $\sim$1.7 keV is seen towards the centre of A2151C(B).  It rises to $>$ 2 keV as one moves further out from the X-ray peak but begins to drop again towards the outskirts of the subclump.  The temperature in regions covering A2151C(F) is lower than that in A2151C(B). The fractional errors in the temperature values are $\sim$6 $-$ 14 per cent. The abundance map is shown in Fig. \ref{fig:fig7}\textcolor{blue}{(b)}. The distribution of metals seems to be fairly symmetric about the X-ray emission peak of A2151C(B).  In the inner regions of A2151C(B), the metallicity is close to or as high as the solar value (Z$_{\odot}$) after taking errors of $\sim$30 per cent in these regions into consideration.  In contrast, the metallicity value in the interior of A2151C(F) is at most $\sim$0.27 Z$_{\odot}$.  The map also reveals that the abundance in the outer regions of both the subclumps is very poor ($\leq$ 0.1 Z$_{\odot}$).   The projected maps of pressure and entropy are shown in Fig. \ref{fig:fig7}\textcolor{blue}{(c)} and Fig. \ref{fig:fig7}\textcolor{blue}{(d)}, respectively.

\section{Discussion}\label{sec4}
We find that the gas in the two groups of A2151C:  A2151C(B) and A2151C(F), has an average temperature of $2.01\pm{0.05}$ keV and $1.17\pm{0.04}$ keV respectively (Table \ref{tab:tab2}).  The temperatures obtained are in agreement  with the \textit{ROSAT} results of \cite{bdb95} but with tighter contraints. The X-ray luminosity of A2151C(B) is $3.03^{+0.02}_{-0.04} \times 10^{43}$ erg s$^{-1}$ (0.4 -- 7.0 keV) while that of A2151C(F) is $ 1.13\pm{0.02} \times 10^{43}$ erg s$^{-1}$ (0.4 -- 7.0 keV).  The X-ray luminosities are $\sim$3 times higher than those reported by \cite{bdb95} due to larger bandwidth and the extent of the gas covered here, though the ratio between the `bright' and the `faint' subclumps is similar.  The temperature and luminosity values obtained for the two subclumps are representative of groups of galaxies \citep{bahcall1999,helsdon2000,rasmussen2007,lovisari2015,zou2016}. The average elemental abundances in A2151C(B) and A2151C(F) are $0.43\pm{0.05}$ and $ 0.13\pm{0.02}$ relative to solar values, respectively (Table \ref{tab:tab2}). The abundance of A2151C(B) is consistent with the value obtained by \citet{bdb95}. The average abundance of A2151C(F) is, however, lower than the \textit{ROSAT} value of $0.32^{+0.15}_{-0.10}$. The different values of the temperature and abundance of the gas in the two subclumps corroborate the optical finding that they are indeed two separate galaxy groups. The total gravitational mass of the groups derived under the assumption of hydrostatic equilibrium is estimated to be $1.60\pm{0.40} \times 10^{13}$ M$_{\odot}$ (within a radius of 304 kpc) for A2151C(B) and $1.44\pm{0.62} \times 10^{13}$ M$_{\odot}$ (within 326.8 kpc) for A2151C(F) (\S3.6). The total mass value obtained for A2151C(B) in this study is found to be consistent with that derived by \citet{huangsarazin96}.The gas mass is derived to be $1.89_{-0.84}^{+1.31} \times 10^{12}$ M$_{\odot}$ (within a radius of 304 kpc) for A2151C(B) and $1.48_{-0.44}^{+0.56} \times 10^{12}$ M$_{\odot}$ (within 326.8 kpc) for A2151C(F) (\S3.5). These values are also typical of galaxy groups \citep{mulchaey2000,lovisari2015} in contrast to galaxy clusters which have a greater extent and possess higher mass, temperature and luminosity \citep{bohringer2002,reiprich2002,sanderson2006}. The $M_{gas\_500}$ and $M_{grav\_500}$ values of A2151C(B) are $1.24_{-0.61}^{+0.97} \times 10^{13}$ M$_{\odot}$ and $9.08\pm{5.24} \times 10^{13}$ M$_{\odot}$ respectively, while these values are $6.11_{-1.97}^{+2.66} \times 10^{12}$ M$_{\odot}$ and $3.01\pm{2.11}\times 10^{13}$ M$_{\odot}$ for A2151C(F). Both $M_{gas\_500}$ and $M_{grav\_500}$ values of the two subclumps are similar to those of galaxy groups \citep{lovisari2015}.

The combined X-ray image of A2151C(B) and A2151C(F) (Fig. \ref{fig:fig1}(a)) shows a region of overlap in X-ray emission  between the two groups. The optical analysis of \citet{agulli2017} indicates that the two groups are located at similar redshifts (radial velocities 10116 km s$^{-1}$ and 10299 km s$^{-1}$ for A2151C(B) and A2151C(F) respectively). Therefore, the possibility of a group-group merger cannot be completely ignored. The thermodynamic maps in \autoref{fig:fig7} do not show an enhancement in temperature, entropy or metallicity in the region between the two groups. The idea of an ongoing group scale merger, thus seems unlikely. We, however, point out that the box size chosen for the spectral analysis in the overlapping region is rather large due to low statistics.  Consequently, a rise in the temperature or entropy (which may be visible in a relatively narrower region in projection where the two groups may be interacting) may have been masked out. Higher resolution spectral analysis with better statistics, e.g., from a  deep exposure with \textit{Chandra}  covering this overlapping region, would be required to shed light on the merger proposition.

A2151C(B) shows evidence of a cool core based on an excess of X-ray emission at the center compared to the single $\beta$ model fitted to the surface brightness profile obtained from the HRI image \citep{huangsarazin96}. The spectral analysis of the 2D projected emission from A2151C(B) presented in this study shows a drop in temperature from 2.31$\pm{0.14}$ keV at radii of 30$-$120 kpc to $1.83_{-0.07}^{+0.08}$ keV in the central 30 kpc region (Fig. \ref{fig:fig3}\textcolor{blue}{(a)} and Table \ref{tab:tab3}) thus confirming the result of \citet{huangsarazin96}. The deprojected spectral analysis gives an even lower value of central temperature (${1.64}_{-0.06}^{+0.08}$ keV; Fig. \ref{fig:fig4}(a) and Table \ref{tab:tab4}). The projected maps of temperature in Fig. \ref{fig:fig6}\textcolor{blue}{(a)} and Fig. \ref{fig:fig7}\textcolor{blue}{(a)}, and the low central entropy seen in Fig. \ref{fig:fig6}\textcolor{blue}{(d)} and Fig. \ref{fig:fig7}\textcolor{blue}{(d)}, indicate the existence of a cool core in A2151C(B).  Furthermore, estimated cooling time of $8.05_{-0.76}^{+0.84} \times 10^{8} \text{ yr}$ (significantly smaller than the Hubble time) in the central 15 arcsec region of A2151C(B) (\S3.7) reinforces the presence of a cool core in the subclump. A2151C(F), in contrast, does not show significant evidence of containing a cool core. Although the projected spectral analysis indicates a gradual decrease in temperature from $1.32\pm{0.10}$ keV at radii of 76$-$114 kpc to $1.07\pm{0.04}$ keV in the central 38 kpc region of A2151C(F) (Fig. \ref{fig:fig3}\textcolor{blue}{(d)} and Table \ref{tab:tab3}), this trend is not strong enough to classify this subclump as a definitive cool-core group. The deprojected temperature profile of A2151C(F) (Fig. \ref{fig:fig4}\textcolor{blue}{(c)}) does not show a drop in temperature towards the central region. A2151C(F), therefore, comes across as a weak-cool-core or non-cool-core group. Heating from radio sources at the centres of groups and clusters is known to inhibit the central gas from cooling \citep{matthews2003,mcnamara2007,gitti2012,ming2012} and the presence of the wide-angle tailed (WAT) galaxy NGC 6047 in the centre of A2151C(F) \citep{huangsarazin96} with a total radio luminosity of $7.5\times 10^{40} \text{ erg s}^{-1}$ (integrated from 10 MHz to 10 GHz; \citet{feretti1988}) may be responsible for arresting the development of a cool core within this group.

The presence of significant clumping of gas seen in A2151 -- the bimodality in A2151C seen in Fig. \ref{fig:fig1} in addition to the gas clumps seen towards further east (A2151E) and north (A2151N) in the \textit{ROSAT} PSPC image (fig. 1 of \citet{bdb95}) -- contrasts it with relaxed and evolved clusters like Coma \citep{neumann2003}. The clumped and irregular X-ray morphology observed in A2151C is similar to that observed in the nearby cluster Abell 76 which is believed to be in the early phase of cluster formation \citep{ota2013}. Similar clumpiness is also found in A2147 \citep{jones1979,reichert1981} which along with A2151 and A2152 forms the Hercules supercluster, and is also considered to be in its early evolutionary phase. Moreover, the high spiral fraction of galaxies in A2151 ($\gtrsim$50 per cent; \citet{gioandhaynes85,bdb95,maccagni95}) is also indicative of the relatively unevolved state of the cluster as a whole.

The bimodal structure in A2151C seen in X-rays \citep{magri88,bdb95,huangsarazin96} as well as in the galaxy distribution \citep{huangsarazin96,agulli2017}, suggests that A2151C(B) and A2151C(F) are two physically distinct entities.  On one hand, A2151C(B) possesses a cool core along with a fairly regular X-ray emission reflecting its relaxed and virialised nature. On the other hand, the irregular X-ray morphology of A2151C(F) together with its low X-ray luminosity, temperature and abundance are indications that this subclump is in the process of formation. The presence of several compact galaxy groups (CGGs) identified within A2151C(F) (Fig. \ref{fig:fig1}\textcolor{blue}{(b)}) lends further support to this idea.   We notice an increase in temperature (Fig. \ref{fig:fig7}(a)) and entropy (Fig. \ref{fig:fig7}\textcolor{blue}{(d)}) in the eastern region within A2151C(F). The region overlaps with the CGG SDSSCGB 4240 (Fig. \ref{fig:fig1}\textcolor{blue}{(b)}) in the outskirts of A2151C(F), which is probably merging with the subclump and leading to the observed enhancement. The disagreement between the $\beta$ value obtained from fitting the gas density profile ($\beta_{fit}=0.28\pm{0.02}$) and that derived using \autoref{eq4} ($\beta\sim$3; \S3.5), further suggests the dynamically unrelaxed and evolving status of A2151C(F).   Although galaxy groups are known to have $\beta_{fit} \lesssim$ 0.5 \citep{helsdon2000,mulchaey2000}, the value obtained for A2151C(F) is unusually low.

\citet{bdb95} did not attribute the bimodality in X-ray emission in A2151C to correspond to two separate features in their galaxy density map (fig. 2 of their paper).  Instead, they reported a single galaxy density peak associated with A2151C.  Based on this and the presence of the WAT source within A2151C(F) (often associated with clusters mergers; see references in \citet{bdb95}), they postulated that the gas identified with A2151C(F) is the ram-pressure stripped part of A2151E (the gas clump towards further east; not covered in this study but seen in the \textit{ROSAT} PSPC image) which got trapped within the deeper potential well of A2151C(B), when a merger event between A2151C(B) and A2151E took place.   Our analysis presents a different scenario. The X-ray emission in A2151C(F) is now seen to coincide with a definite structure in the spatial distribution of galaxies fig 2 of \citet{huangsarazin96} and fig.5 of \citet{agulli2017}). Besides, A2151C(B) appears to be fairly symmetric and undisturbed, and also contains a cool core which seems to be intact even after the supposed occurrence of a merger event. The radio jets associated with NGC 6047 extend $\sim$52 arcsec or $\sim$40 kpc across in a N-S direction and form a small-scale structure (Fig. 9 of \citet{huangsarazin96}). Their extent is comparable to the optical diameter of the galaxy ($\sim$81 arcsec (SDSS r-bandpass; ref: NED) or $\sim$61 kpc). The interstellar medium rather than a merger activity may therefore play a non-negligible role in shaping the tails, as seen in galaxy NGC 4874 within the Coma cluster \citep{feretti1985}, in contrast with large-scale WATs ($\gtrsim$ 200 kpc) which are often thought to be created in cluster mergers \citep{sakelliou2000}. Nonetheless, if the hypothesis of \citet{bdb95} is assumed to be true, it may be possible that the merger between A2151C(B) and A2151E was either a minor one or it took place at a time when A2151C(B) had already developed a well defined cool core (simulations show that major mergers tend to disrupt cool cores (\citet{ritchie2002,zuhone2011}), more so if the mergers take place in the early evolutionary stage of the cluster (\citet{burns2008,henning2009})). However, to establish the correctness of the tentative post-merger scenario proposed by \citet{bdb95} or rule it out, deep exposures with \textit{Chandra} as well as radio observations targeting the entire A2151 field (including both A2151C and A2151E), revealing the presence/absence of shocks, are required.

\section{Conclusions}\label{sec5}
Our detailed X-ray spectral study of two subclumps of diffuse X-ray emission in A2151C -- A2151C(B) and A2151C(F) -- is based on a  broader energy band X-ray data obtained from \textit{XMM-Newton} and \textit{Chandra} (\autoref{fig:fig1}) observations having much better signal-to-noise  ratio than reported earlier using the \textit{ROSAT}, and provides better constraints on the thermodynamic properties of the hot gas.  Our results strengthen the identification of these subclumps  with two distinct galaxy groups seen in the optical data recently. We give a detailed description of the X-ray gas in A2151C(F) regarding it as a separate galaxy group.  We have tested for the presence of cool cores within the two groups, and provided estimates for the total mass and gas mass in them. We have presented the 2D projected maps of metallicity and thermodynamic (TD) properties of the gas in A2151C and searched for signatures of small scale mergers within the ICM.  Since A2151C(B) and A2151C(F) are located at similar redshifts and overlap in projection (seen in their X-ray image), we have investigated the possibility of an ongoing interaction between the two groups.
The main results of our study are as follows:
\begin{itemize}\setlength\itemsep{1em}
\item X-ray emission from the western subclump, A2151C(B), of the central subcluster of Hercules has $L_{X (0.4-7.0)} = 3.03^{+0.02}_{-0.04} \times 10^{43}$ erg s$^{-1}$,  while the eastern component, A2151C(F), is fainter with $L_{X (0.4-7.0)} = 1.13\pm{0.02} \times10^{43}$ erg s$^{-1}$. 
\item A2151C(B) displays a fairly regular X-ray morphology, whereas A2151C(F) is quite irregular. Several compact groups of galaxies are also seen within A2151C(F).
\item Low average temperature, $ 2.01\pm{0.05}$ keV for A2151C(B) and $1.17\pm{0.04}$ keV for A2151C(F), representative of galaxy groups rather than clusters is confirmed.
\item A2151C(B)  has higher elemental abundance of $0.43\pm{0.05}$ Z$_{\odot}$ than A2151C(F) which has an abundance of only $0.13\pm{0.02}$ Z$_{\odot}$, which is lower than the
value obtained with the \textit{ROSAT} (\S1).
\item Low temperature (${1.55}\pm{0.07}$ keV) in the central 15 arcsec region in addition to a short cooling time of $\sim$0.81 Gyr confirms the presence of a cool core in A2151C(B).  There is no convincing evidence for a cool core in A2151C(F), however.
\item Mass of hot gas is estimated to be  $ 1.89_{-0.84}^{+1.31} \times 10^{12}$ M$_{\odot}$ (within a radius of 304 kpc) for A2151C(B) and $1.48_{-0.44}^{+0.56} \times 10^{12}$ M$_{\odot}$ (within 326.8 kpc) for A2151C(F). 
\item The total gravitational mass of A2151C(B) within a radius of 304 kpc is estimated to be $1.60\pm{0.40}\times10^{13}$ M$_{\odot}$ while for A2151C(F) it is estimated to be $1.44\pm{0.62} \times 10^{13}$ M$_{\odot}$ within a radius of 326.8 kpc. The total gravitational mass and mass of the hot gas in A2151C(F) have been estimated for the first time here.
\item{The $M_{gas\_500}$ and $M_{grav\_500}$ values of A2151C(B) are $1.24_{-0.61}^{+0.97} \times 10^{13}$ M$_{\odot}$ and $9.08\pm{5.24} \times 10^{13}$ M$_{\odot}$ respectively, while these values are $6.11_{-1.97}^{+2.66} \times 10^{12}$ M$_{\odot}$ and $3.01\pm{2.11}\times 10^{13}$ M$_{\odot}$ for A2151C(F).}
\item We do not find an enhancement of temperature or entropy in the overlapping region between A2151C(B) and A2151C(F), thus making the possibility of an ongoing merger between the two groups seem unlikely.
\item A2151C(B) is identified as a rich and relaxed group, while A2151C(F) appears to be in the process of formation.
\end{itemize}
\vspace{4mm}\textbf{ACKNOWLEDGEMENTS}\vspace{2.5mm}
\\The X-ray data used in this research were obtained from the High Energy Astrophysics Science Archive Research Center (HEASARC), maintained by NASA's  Goddard Space Flight Center. We have used X-ray observations obtained with the XMM-Newton telescope, a science mission of the European Space Agency (ESA), and the Chandra X-ray observatory, managed by NASA's Marshall Center. We have also used optical imaging data from data release 12 of the Sloan Digital Sky Survey-III (\url{http://www.sdss3.org/}). Funding for SDSS-III has been provided by the Alfred P. Sloan Foundation, the Participating Institutions, the National Science Foundation, and the U.S. Department of Energy Office of Science. This research has made use of SAOImageDS9, developed by Smithsonian Astrophysical Observatory, and the HEASoft FTOOLS (\url{http://heasarc.gsfc.nasa.gov/ftools}). This research has also made use of the NASA/IPAC Extragalactic Database (NED), operated by the Jet Propulsion Laboratory, California Institute of Technology, under contract with the National Aeronautics and Space Administration, and the SIMBAD database, operated at CDS, Strasbourg, France.
We thank the Chandra and XMM helpdesk for their assistance with the X-ray data analysis. We are grateful to Keith Arnaud for helping us to gain a better understanding of the working and implementation of XSPEC models. We thank the reviewer for the useful comments that greatly helped in improving the content of the paper.
\\\\\textbf{\\\\Data availability}\vspace{2.5mm}
\\This article has made use of archival data from observations with the European Photon Imaging Camera (EPIC) onboard the XMM-Newton observatory and the Advanced CCD Imaging Spectrometer (ACIS) onboard the Chandra X-ray Observatory (CXO). All X-ray data were obtained from the HEASARC archive and are publicly accessible via \url{https://heasarc.gsfc.nasa.gov/cgi-bin/W3Browse/w3browse.pl}. 
\bibliographystyle{mnras}
\bibliography{hercules} 
\bsp
\label{lastpage}
\end{document}